\documentclass[onecolumn]{revtex4-2}
\usepackage[utf8]{inputenc}
\usepackage{tabularx}
\usepackage{array}
\usepackage{bbold}
\usepackage{bm}
\usepackage{hyperref}
\usepackage{amsmath}
\usepackage{graphicx}
\graphicspath{{graphics/}}
\usepackage{xfrac}
\usepackage{ulem}
\usepackage{float}
\usepackage{wrapfig}
\usepackage{caption}
\usepackage[dvipsnames]{xcolor}
\usepackage{mathrsfs,amsmath,amssymb,bm}
\allowdisplaybreaks
\usepackage{comment}

\hypersetup{
	colorlinks=true,
	linkcolor=blue,
	citecolor=blue,			
	urlcolor=blue,			
	filecolor=blue,
	anchorcolor = green,
}

\begin{document}
\title{Diffuselet method for three-dimensional turbulent mixing of a cloudy air filament}
\author{Vladyslav Pushenko$^1$, Simone Scollo$^2$, Patrice Meunier$^2$, Emmanuel Villermaux$^{2,3}$ and J\"org Schumacher$^{1,4}$}
 \affiliation{$^1$Institute of Thermodynamics and Fluid Mechanics, P.O.Box 100565, Technische Universität Ilmenau, D-98684 Ilmenau, Germany\\
                $^2$Institut de Recherche sur les Phénomènes Hors Equilibre (IRPHE), CNRS, Aix-Marseille Universit\'{e}, 13384 Marseille Cedex 13, France\\
                $^3$Institut Universitaire de France,  75005 Paris, France\\
                $^4$Tandon School of Engineering, New York University, New York, NY 11201, USA} 

\date{\today}
 
\begin{abstract}
The mixing properties of vapor content, temperature and particle fields are of paramount importance in cloud turbulence as they pertain to essential processes, such as cloud water droplet evaporation and entrainment. Our study examines the mixing of a single cloudy air (which implies droplet-laden) filament with its clear air environment, a characteristic process at the cloud edge, in two ways. The first consists of three-dimensional combined Euler-Lagrangian direct numerical simulations which describe the scalar supersaturation as an Eulerian field and the individual cloud water droplets as an ensemble of Lagrangian tracers. The second way builds on the recently developed diffuselet method, a kinematic Lagrangian framework that decomposes a scalar filament into a collection of small sections subject to deformation by local stirring and cross-sheet diffusion. The Schmidt number is $Sc=0.7$. The entrainment process causes a deformation of the supersaturated cloud filament in combination with diffusion until the system reaches a well-mixed equilibrium state, which implies for the present configuration that all droplets are evaporated. We compare the time dependence of the mean square and probability density function of the supersaturation field. For the initial period of the mixing process they agree very well; at later stages deviations caused by non-zero mean of the conserved scalar are observed.  For the cases including cloud water droplets, we investigate also the impact of droplet number density and condensation growth response. Turbulence causes deviations from the $d^2$-law similar to recent experiments in sprays. A simulation at a Schmidt number that is by a factor of 100 larger than in clouds improves the agreement between simulation and diffuselet method significantly. The latter result promotes the diffuselet framework as an efficient parametrization for turbulent high-$Sc$ mixing which can reduce the resolution efforts of the viscous-convective range of scalar turbulence.
\end{abstract}
\maketitle

\section{Introduction}

Turbulent mixing and the coupling to droplet microphysics in atmospheric clouds remains one major source of uncertainty for more reliable predictions of the ongoing climate change and the connected global warming \cite{Shaw2003,Bodenschatz2010,Stevens2013,Vogel2022}. It is expected that a  better understanding of these small-scale processes will improve their convective parametrization starting from large-eddy simulations of individual clouds all the way up to global atmospheric circulation models. Water exists in warm clouds in the form of water vapor and cloud water droplets \cite{Rogers1989}. The entrainment of dry clear into moist cloudy air has a significant impact on the statistical properties inside the clouds as well as their lifetime \cite{Baker1984}. This also determines  number density and size distribution of the droplets \cite{Andrejczuk2006,Celani2008,Sardina2015,Kumar2018,Fries2021,Kumar2021,Grabowski2022,Fries2023} and thus eventually the appearance of precipitation \cite{Magaritz2016}. 

Turbulent mixing \cite{Dimotakis2005,Sreenivasan2010,Sreenivasan2019} in general combines two fundamental processes, the stirring of a scalar substance due to advection by the (turbulent) flow and molecular diffusion \cite{Villermaux2019}. A turbulent flow at its smallest scales is characterized by spatially smooth velocity fields; they stretch and fold scalar filaments in a range of decreasing scales that starts around the viscous Kolmogorov length $\eta_K$ \cite{Batchelor1959,Kraichnan1968}. In this way, fluid motion leads to an ongoing aggregation of scalar filaments if the scalar diffusion is sufficiently small (which is quantified in Sec. II by the dimensionless Schmidt number). A specific form of the probability density function of the scalar concentration field in the form of a gamma distribution follows from this picture \cite{Villermaux2003,Meunier2003,Meunier2010}. At the smallest spatial scales in the mixing problem, molecular diffusion of the scalar will always dominate the dynamics. This is the beginning of the inertial-diffusive and viscous-diffusive scale range for Schmidt (or Prandtl) numbers $Sc\ll 1$ and $Sc\gg 1$, respectively. 

The primary goal of the present work is to investigate whether a kinematic Lagrangian framework that incorporates these elementary building blocks of the turbulent mixing process -- the diffuselet model \cite{Meunier2022} -- is applicable to the entrainment at the edge of a turbulent cloud. Our model system is a thin initially slab-like filament of cloudy supersaturated air in a clear air environment at the same turbulent kinetic energy level, similar to previous numerical studies \cite{Kumar2012,Kumar2013,Kumar2018,Kumar2014,Goetzfried2017}. To this end, we will proceed in two subsequent steps. First, we consider the passive scalar mixing model of the supersaturation field \cite{Celani2005,Fries2021,Pushenko2024} and investigate how the diffuselet model compares to fully resolved direct numerical simulations (DNS) of three-dimensional scalar turbulence in a periodic box, see e.g. \cite{Pushenko2024,Schumacher2007}. Two cases are studied here, one at a Schmidt number of $Sc=0.7$, which corresponds to atmospheric conditions, and one at a hundred times larger value of $Sc=70$ which develops a viscous-convective range for scales $r\lesssim \eta_K$. The latter case is added even though this high Schmidt number cannot be related to cloud turbulence. The early period of the mixing process is well described by the diffuselet model; at later times we will observe growing deviations and discuss the reasons.

Secondly, we add an initially monodisperse ensemble of cloud water droplets to the scalar filament and numerically solve the cloud model \cite{Pushenko2024}. We investigate how strongly the deviations from the diffuselet model grow with time in comparison to the pure scalar field case. Therefore, we enhance the droplet number density beyond realistic values of $n_d\sim 10^2$ cm$^{-3}$ \cite{Shaw2003,Rogers1989}. Furthermore, we increase the cloud water droplet response to super-/subsaturation by an increase of the diffusion constant $K$ in the droplet growth equation \cite{Kumar2012}. Our results are also compared with dense spray mixing experiments \cite{Rivas2016}. In summary, we find that the additional impact of the droplets on the presently chosen configuration remains small, such that the diffuselet method in its present form is even applicable (in the same limits as for the passive scalar case) to a two-phase mixing process. A major scope of this work is to explore the potential, but also the limitations of the diffuselet model in absence of a viscous-convective scale range of the scalar field for the transport in a three-dimensional complex Navier-Stokes flow.

The outline of the manuscript is as follows. In Sec. II, we describe the equations, parameters and physically relevant scales and present a brief summary of the central ideas of the diffuselet model. Sec. III presents all results, starting with a test of the analytical model, via the pure scalar case to the filament mixing with droplets. The manuscript ends with our conclusion and an outlook in Sec. IV. Technical details on the simulations are summarized in an appendix.   

\section{Methods}
\subsection{Eulerian fields and direct numerical simulations of turbulent mixing}
The present cloud mixing model is based on three fields: the vapor mixing ratio $q_v({\bm x},t)$, the liquid water mixing ratio $q_l({\bm x},t)$, and the temperature field $T({\bm x},t)$, see ref. \cite{Pushenko2024} for further details. The fallout of rain is not included, such that $q_v+q_l={\rm const.}$ The modeling of the liquid water mixing ratio $q_l$ in the Lagrangian framework will be detailed in Sec. II B. The Eulerian fields $q_v$ and $T$ are combined in the scalar supersaturation field $s({\bm x},t)$ which is approximated as a {\it passive scalar} that determines the diffusion growth of the cloud water droplets. It is given by 
\begin{equation}
    s({\bm x},t) = \frac{q_v({\bm x},t)}{q_{\rm vs}(T({\bm x},t))}-1\,,
\end{equation}
where $q_{\rm vs}$ is the saturated vapor mixing ratio that depends on temperature $T$, determined by the Clausius-Clapeyron equation \cite{Rogers1989}. For convenience, we  rescale supersaturation $s$ to the scalar field $c$ in the following way, 
\begin{equation}
    0\le c({\bm x},t) = \frac{s({\bm x},t)+1}{s_{\rm max}+1} \le 1\,.
    \label{eq:scalar_rescale}
\end{equation}
The transport of $c$ is determined by
\begin{equation}
    \frac{\partial c}{\partial t}+({\bm u}\cdot{\bm \nabla})c=D \nabla^2 c\,,
    \label{eq:transoprt_c}
\end{equation}
where $D$ is the scalar diffusivity and ${\bm u}$ the advecting (turbulent) velocity field. The dynamics of the latter field is given by the incompressible Navier-Stokes equations,
\begin{align}
\label{eq:nse}
{\bm \nabla} \cdot {\bm u} = 0\quad\quad \mbox{and}\quad\quad
\frac{\partial {\bm u}}{\partial t} + ({\bm u} \cdot {\bm \nabla}) {\bm u} = - \frac{1}{\rho_0}{\bm \nabla} p + \nu{\bm \nabla}^2 {\bm u} + {\bm f}\,.    
\end{align}
Here, $p({\bm x},t)$ is the pressure field, $\rho_0$ the constant mass density, and $\nu$ the kinematic viscosity. Furthermore, ${\bm f}({\bm x},t)$ is a large-scale volume forcing which injects turbulent kinetic energy into the fluid flow at a fixed rate, see further below in this section for details. The dimensionless form of \eqref{eq:transoprt_c}--\eqref{eq:nse} is obtained by defining characteristic values of all physical quantities. This is the length of the cubic simulation domain $L$, the reference scalar amplitude $c_0$, and the root-mean-square value of the velocity field $U$, which leads to
\begin{align}
\label{eq:sys1}
{\bm \nabla} \cdot {\bm u} &= 0\,,\\
\label{eq:sys2}
\frac{\partial {\bm u}}{\partial t} + ({\bm u} \cdot {\bm \nabla}) {\bm u} &= -{\bm \nabla} p + \frac{1}{Re}{\bm \nabla}^2 {\bm u} + {\bm f}\,,\\
\label{eq:sys3}
\frac{\partial c}{\partial t }+({\bm u}\cdot {\bm \nabla}) c  &=  \frac{1}{Re\, Sc}  \nabla^2 c\,.
\end{align}
Here, $Re={UL}/\nu$ is the large-scale Reynolds number and $Sc=\nu/D$ is the Schmidt number. The Schmidt number $Sc$ relates consequently momentum to scalar diffusion.

In three-dimensional turbulent flows, the turbulent kinetic energy, which is transferred from large to small scales, is converted into heat starting at the Kolmogorov scale $\eta_K=(\nu^3/\langle\epsilon\rangle)^{1/4}$ with the mean rate of kinetic energy dissipation, $\langle\epsilon\rangle$ \cite{Kolmogorov1941}. This cascade picture can be applied to the scalar variance cascade depending on $Sc$ \cite{Sreenivasan2010}. For $Sc\leq 1$, the scalar variance cascades down to smaller scales in the inertial-convective range, which is bounded from below by the Corrsin length $\eta_C=(D^3/\langle \epsilon\rangle)^{1/4} \gtrsim \eta_K$, as suggested by Obukhov \cite{Obukhov1949} and Corrsin \cite{Corrsin1951} independently. For $Sc\ge 1$, the scalar variance cascades down the inertial range to $\eta_K$. For even smaller scales $r$, the viscous-convective range is established up to the Batchelor scale $\eta_B=\eta_K/\sqrt{Sc}=\sqrt{D/\gamma}$ with the mean stretching rate $\gamma$ \cite{Batchelor1959,Kraichnan1968}. In this range between $\eta_K \gtrsim r \gtrsim \eta_B$, the scalar is stirred by a spatially smooth flow, which can be considered as a chaotic mixing regime \cite{Ottino1990,Meunier2003}. It becomes clear now, that the DNS at $Sc=0.7$ is assigned to Corrsin-Obukhov regime of passive scalar turbulence, the one at $Sc=70$ to the Batchelor-Kraichnan regime.      

The DNS apply the pseudospectral method. Turbulent velocity and passive scalar field, ${\bm u}({\bm x},t)$ and $c({\bm x},t)$, are expanded in Fourier modes with respect to each of the three space directions relying on volumetric fast Fourier transformations of the software package P3DFFT \cite{Pekurovsky2012}. Time stepping is done with a second-order predictor-corrector scheme \cite{Yeung1988}, see also ref. \cite{Pushenko2024}. Further details on the required spectral resolution and the resulting Reynolds numbers will be listed in tables \ref{table:turb_cases} and \ref{table:dr_cases}. The computational mesh is uniform in all three space directions with a mesh size $\Delta x$. 

The volume forcing ${\bm f}$ is applied in Fourier space for a selected set of low-wavenumber Fourier modes, following the scheme introduced in ref. \cite{Schumacher2007}. More specifically, the forcing acts on a small subset ${\cal K}$ of wavevectors
\begin{equation}
    {\cal K}=\left\{k_j^f=\frac{2\pi}{L}(\pm1,\pm1,\pm2)\;\;\mbox{plus permutations} \right\} \,,
\end{equation}
ensuring an isotropic kinetic energy injection at the largest scales. The forcing is given by
\begin{equation}
    f_i (k_j,t)=\epsilon_{\rm inj}\,\dfrac{\widehat{u}_i(k_j,t)}{\sum_{k_j^f\in\mathcal{K}}\left|\widehat{u}_i(k_j^f,t)\right|^2}\,\delta_{k_jk_j^f}\,.
    \label{eq:forcing}
\end{equation}
In statistically steady state, this balances the mean energy dissipation rate, such that $\langle\epsilon\rangle = \epsilon_{\rm inj}$, resulting in a turbulent velocity field with a Kolmogorov length scale which can be prescribed. The kinetic energy injection rate, $\epsilon_{\rm inj}=\langle {\bm u}\cdot{\bm f}\rangle$, is chosen such that $\eta_K = 1$ mm, which is typical of cloud turbulence. So far, we have discussed the treatment of Eulerian fields (velocity, passive scalar). The Lagrangian part follows now in subsection II B.

\subsection{Lagrangian diffuselet framework of turbulent mixing}
\begin{figure}
    \centering
    \includegraphics[width=0.85\linewidth, trim={0 3cm 0 0},clip]{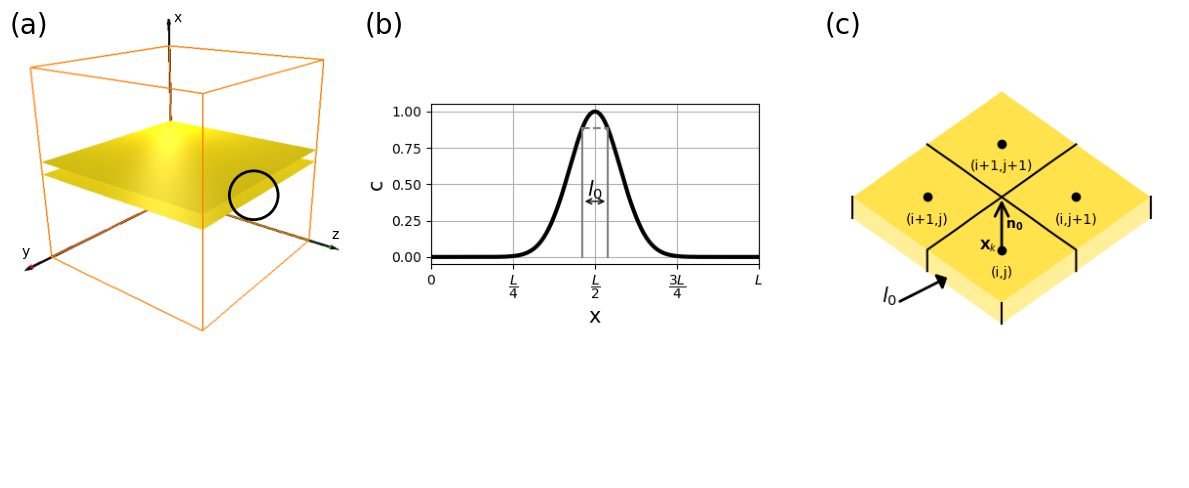}
    \caption{The sheet configuration of the initial concentration field $c({\bm x},0)$. (a) The sheet is placed in the triply periodic simulation domain $V=L^3$ parallel to the $y$-$z$-plane and centered at $x=L/2$. (b) The cross-sheet concentration profile is a Gaussian, see eq. \eqref{eq:DefInitialProfile}, characterized by a thickness $\ell_0$ in $x$-direction. The thicknesses in (a) and (b) are magnified for better visualization. (c) Decomposition of the sheet into diffuselets. Center points ${\bm X}_k(t=0)$ are indicated as black dots as well as the normal vector ${\bm n}_0$. Index $k$ ranges from 1 to $N_{\rm diff}$.}
    \label{fig:init_profile}
\end{figure}
In accordance with ref. \cite{Meunier2022}, a scalar sheet is divided into $N_{\rm diff}$ finite-size elements, as illustrated in Fig. \ref{fig:init_profile}. The sheet is very thin, such that the well-mixed equilibrium will correspond to a mean vapor content below the saturation level. This implies that all droplets are evaporated. For simplicity, we select a Gaussian transversal profile for the initial scalar field,
\begin{equation}
    c(x,t=0) = \exp\left(-\frac{(x-L/2)^2}{\ell_0^2}\right)\,,
 \label{eq:DefInitialProfile}
\end{equation}
with a sheet thickness $\ell_0$. This profile translates to the same Gaussian shape for the supersaturation field $s$. The latter field is positive, i.e., supersaturated in a cross section of approximately $\ell_0/2$ with $s_{\rm max} =0.05$ in the sheet center, see eq. \eqref{eq:scalar_rescale}. Each finite-size element is associated with a Lagrangian tracer, whose coordinates correspond to the coordinates of the geometrical center of the element, see Fig. \ref{fig:init_profile}(c). Lagrangian tracers are seeded in the center plane of the volume with initial positions $\{{\bm X}_i^0\}_{i=1,..,N_{\rm diff}}$. Their trajectories $\{{\bm X}_i(t)\}_{i=1,..,N_{\rm diff}}$ are given by
\begin{equation}
\frac{d{\bm X}_i (t)}{dt} = {\bm u} ({\bm X}_i,t), \quad i=1,...,N_{\rm diff} \quad\mbox{with}\quad 
{\bm X}_i(t=0) = {\bm X}_i^0\,.
\label{eq:lagr_trac}
\end{equation}
These elements are referred as {\it diffuselets}, and the transport equation (\ref{eq:transoprt_c}) can be solved for each diffuselet individually. The resulting data will then be employed in the reconstruction of the global statistical properties of the scalar field. Time integration of the particle (or diffuselet) trajectories is performed with the same second-order predictor-corrector method as the Eulerian fields. The interpolation of the velocity field to each diffuselet position is tri-linear.

The concentration field of each diffuselet can be calculated using the Ranz coordinate transformation \cite{Ranz1979}. The maximal concentration $c_i$ of the scalar field inside the diffuselet $i$ with the surface $\delta A_i$ and thickness $\ell_i$ is simply deduced from a new dimensionless time $\tau_i$ by $c_i = 1/\sqrt{\tau_i}$. This dimensionless time is determined numerically by integrating ($D$ and $\ell_i$ carry their physical dimension)
\begin{equation}
    \frac{d \tau_i}{dt} = \frac{4 D}{\ell_i^2}, \quad\mbox{with}\quad \tau_i(t=0) = 1.
\label{ranztime}    
\end{equation}
The advection-diffusion equation (\ref{eq:transoprt_c}) is simplified to pure diffusion equation in the new coordinates $(\xi,\tau)$, which is given by 
\begin{equation}
    \frac{\partial c_i}{\partial \tau} = \frac{1}{4} \frac{\partial^2 c_i}{\partial \xi^2}\,.
\end{equation}
Here, $\xi=n_i/\ell_i$ is dimensionless coordinate normal to diffuselet surface. Each diffuselet is characterized by two key parameters: (1) the initial orientation, designated as ${\bm n}_0$, and (2) the initial surface vector, $\delta {\bm A}_0 = {\bm n}_0\,\delta A_0$. All elements have an identical initial surface $\delta A_0\sim \eta_K^2$. Under the influence of the local velocity gradients the surface elements are subject to stretching and compression. A local stretching factor is defined as $\rho_i = \delta A_i/\delta A_0$. Incompressibility of the velocity field requires $\delta A_i(t) \ell_i(t) = \delta A_0 \ell_0$ for the stretching history with $\delta A_i(t)$ and $\ell_i(t)$ being the surface and thickness of $i$-th diffuselet at time $t$ initiating from $\delta A_0$ and $\ell_0$. More specifically, the kinematic stretching dynamics is determined by the velocity gradient tensor along the Lagrangian tracer track ${\bm X}_i(t)$ that was initiated in the center of diffuselet $i$,  
\begin{equation}
    \frac{d\, \delta {\bm A}_i}{dt}=-({\bm \nabla u})^T[{\bm X}_i(t),t] \ \delta {\bm A}_i.
    \label{eq:ai}
\end{equation}
It is actually more powerful to define a surface dispersion tensor $\bm{\mathcal{L}}_i$ for each tracer, which is solution of 
\begin{equation}
    		\frac{d\bm{\mathcal{L}}_i [t]}{dt} =-({\bm \nabla u})^T[{\bm X}_i(t),t] \ \bm{\mathcal{L}}_i[t] \quad\mbox{with}\quad \bm{\mathcal{L}}_i[0]=\mathbb{1}\,.		
\end{equation}
Multiplying this equation on the right by $\delta {\bm A}_0$ indicates that $\bm{\mathcal{L}}_i(t) \delta {\bm A}_0 $ is solution of \eqref{eq:ai}. The surface vector is thus given by
\begin{equation}
    \delta {\bm A}_i(t) =\bm{\mathcal{L}}_i(t)\delta {\bm A}_0\,,
\end{equation}
whatever the initial surface vector $\delta {\bm A}_0$.
The stretching factor of the surface element $\rho_i(t) =  (\delta {\bm A}_i^T \delta {\bm A}_i )^{1/2} / \delta A_0 $  simply writes
\begin{equation}
    \rho_i(t) = \sqrt{{\bm n}_{0,i}^T \mathbf{\mathcal{L}}_i(t)^T \mathbf{\mathcal{L}}_i(t) {\bm n}_{0,i}}\,.
    \label{eq:stretching_fac}
\end{equation}
It depends only on the initial orientation ${\bm n}_{0,i}$ of the diffuselet. It is possible to average $\rho_i$ over all initial orientations in order to accelerate the convergence of the statistics \cite{Meunier2022}.  However, in this paper, we chose to fix ${\bm n}_{0,i}$ along the $x$ direction to be closer to the DNS results. The difference between the two results was found to be very small.

Using the incompressibility, the evolution of the diffuselet thickness $\ell_i(t)$ can be written as
\begin{equation}
		\ell_i(t)=\frac{\ell_0\, \delta A_0}{\delta A_i(t)}=\frac{\ell_0}{\sqrt{{\bm n}_{0,i}^T  \mathbf{\mathcal{L}}_i(t)^T \mathbf{\mathcal{L}}_i(t) {\bm n}_{0,i}}}\,.
\label{eq:s_eq}
\end{equation}
It is then possible to generalize the dimensionless time to a dimensionless tensor, see again eq. \eqref{ranztime}, 
\begin{equation}
\tau_i(t)=1+4D\int_0^t \frac{1}{\ell_i^2(t^{\prime})} dt^{\prime} \longrightarrow
	\bm{\mathcal{T}}_i=\mathbb{1}+\frac{4D}{\ell_0^2}\int_{0}^{t}\left(\bm{\mathcal{L}}_i^T\bm{\mathcal{L}}_i\right)[{\bm X}_i(t'),t']\,dt'\,.
\end{equation}
from which the dimensionless time is obtained as
\begin{equation}
    \tau_i(t) = {\bm n}_{0,i}^T \ \bm{\mathcal{T}}_i(t)  \ {\bm n}_{0,i}
\end{equation}
The maximal concentration of each diffuselet is then equal to $c_i=1/\sqrt{\tau_i}$. The mean square of each diffuselet is to a good approximation the product of the squared maximal concentration $c_i^2$ and the volume of the diffuselet at time $t$, given by $\delta A_0\ell_0/\ell_i\times \ell_i\sqrt{\tau_i}=\delta A_0\ell_0\sqrt{\tau_i}$ and multiplied by a factor of $\sqrt{\pi/2}$ due to the Gaussian profile. This leads to
\begin{equation}
    \langle c_i^2(t)\rangle \approx \frac{\sqrt{\pi}\,\delta A_0\,\ell_0}{\sqrt{2 \tau_i(t)}}
    \label{eq:var_ind_diff}\,.
\end{equation} 
Note that $\ell_i\sqrt{\tau_i(t)}$ is the diffusive thickness in \eqref{eq:var_ind_diff}. The total mean square of all diffuselets is consequently given by sum of \eqref{eq:var_ind_diff} over the entire diffuselet ensemble,
\begin{equation}
	\langle c^2(t)\rangle=\sum_{i=1}^{N_{
    \rm diff
    }} \langle c_i^2(t)\rangle \approx \sum_{i=1}^{N_{\rm diff}}\frac{\sqrt{\pi}\,\delta A_0\,\ell_0}{\sqrt{2 \tau_i(t)}}\,. 
    \label{eq:diff_var}
\end{equation}
Following a similar approach as for the mean square, each diffuselet contributes to the total concentration PDF at the level $c=c_i$ with a weight given by the diffuselet volume $\delta A_0\ell_0/c_i$. The PDF of maximal concentration is then approximately obtained by $\sqrt{\tau_i/\pi}$ for a top-hat profile. A subsequent convolution with the Gaussian profile (which is used here), yields an  additional factor of $1/(c\sqrt{-\log(c\sqrt{\tau_i})}.$ Thus the PDF amplitude at concentration $c$, denoted as $P(c)$, is calculated as the sum over all diffuselets.
\begin{equation}
	P(c)\approx \sum_{i=1}^{N_{\rm diff}}   \frac{ \sqrt{\tau_i/\pi}}{c\sqrt{-\log(c\sqrt{\tau_i})}}.
    \label{eq:pdf_final}
\end{equation}
Equations \eqref{eq:diff_var} and \eqref{eq:pdf_final} will be numerically evaluated with respect to time and compared with the DNS results. More details on the mathematical formulation of the mathematical framework (including the diagonalization of $\bm{\mathcal{L}}_i^T\bm{\mathcal{L}}_i$) can be found in ref. \cite{Meunier2022}. In the appendix, we also report resolution studies, where differently fine decompositions of the filament into diffuselets have been compared to each other. This was done to confirm that mean squared scalar  and scalar PDF remain unaffected for different resolutions. Also in appendix A, we have validated the diffuselet model with an analytical flow example.

\section{Results}    
 
\subsection{Passive scalar mixing in three-dimensional homogeneous isotropic flow}
\subsubsection{Stretching ratio statistics in turbulent flow}
The turbulent mixing in a homogeneous isotropic three-dimensional velocity field adds significant complexity to the problem at hand, caused by the continuum of turbulent eddy scales, which are subject to vortex stretching and other vortex-vortex interactions, see e.g. \cite{Ashurst1987,Girimaji1990,Zinchenko2024}. The dynamics of this turbulent flow requires DNS that solve numerically the incompressible Navier-Stokes equations together with the passive scalar advection-diffusion equation and the Lagrangian diffuselets.

\begin{figure}[t]
\centering
\includegraphics[width=0.8\linewidth]{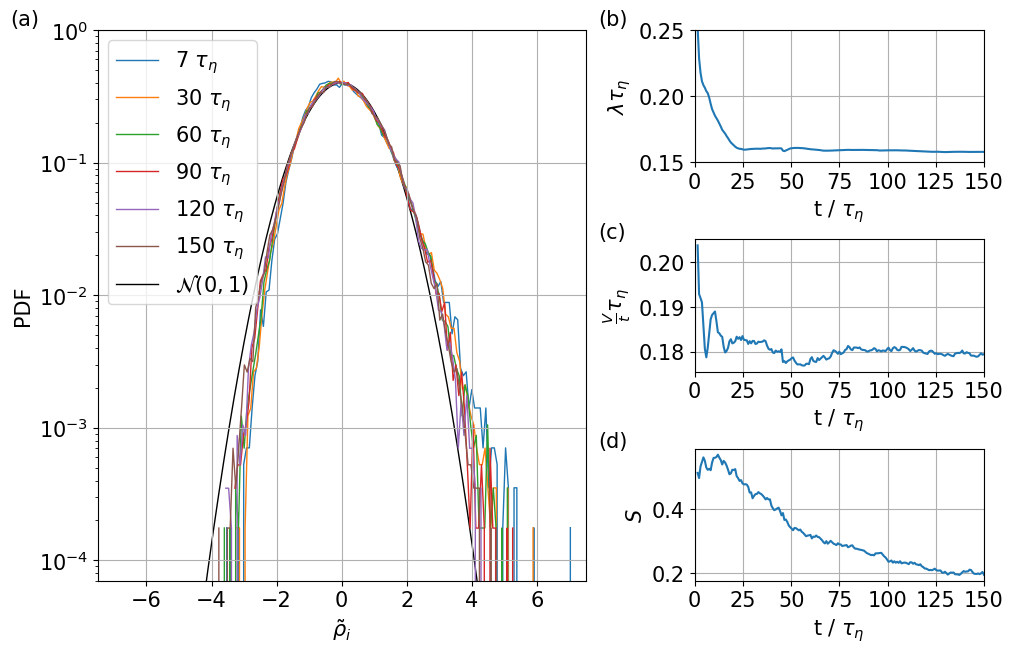}
\caption{Statistics of the surface stretching ratios $\rho_i=\delta A_i(t)/\delta A_0$. (a) PDF of the normalized stretching $\tilde{\rho}_i=\left(\log \rho_i-\langle\log \rho_i\rangle\right)/\sigma(\log\rho_i)$  for different moments of time. The black line represents the normal distribution $\mathcal{N}(0,1)$. (b) Mean finite-time Lyapunov exponent $\lambda = \langle\log\rho_i\rangle/t$ versus time. (c) Temporal evolution of the time-rescaled mean square $V/t$ of $\log\rho_i$, as defined by \eqref{eq:VarianceStretching}. (d) Temporal evolution of the skewness of the PDF of $\log \rho_i$, as defined by \eqref{eq:Skewness}.}
\label{fig:pdf_rho_and_moments}
\end{figure}
    
The statistics of the stretching factors $\rho_i$, see eq. \eqref{eq:stretching_fac}, is shown in Fig. \ref{fig:pdf_rho_and_moments}. We plot the PDF of $\tilde{\rho}_i=\left(\log \rho_i-\langle\log \rho_i\rangle\right)/\sigma(\log\rho_i)$ in panel (a) of the figure. The evolution is similar to that in the seminal work by Girimaji and Pope \cite{Girimaji1990}. Their numerical analysis was conducted at similar, but not exactly the same conditions in the homogeneous isotropic flow. The PDFs are very close to Gaussian, except for a small positive skewness that decreases with time. 

In panels (b)--(d) of the same figure, we list the first three moments of $\log \rho_i$ versus time. The first moment corresponds to the mean Lyapunov exponent 
\begin{equation}
    \lambda=\frac{\langle \log \rho_i \rangle }{t}.
    \label{eq:MeanLyapunovExponent}
\end{equation}
It is initially large and then decreases to a value close to $\lambda=0.159 /\tau_\eta$, where $\tau_\eta=0.07 \: \mathrm{s}$ the Kolmogorov time-scale. This is in fair agreement with  Girimaji and Pope \cite{Girimaji1990} who found a value $\lambda= 0.165 /\tau_\eta $ and who explained the initial decay of the stretching rate by the alignment of the surfaces with the shear. The second central moment is 
\begin{equation}
    V=\langle (\log \rho_i - \lambda t )^2 \rangle\,.
    \label{eq:VarianceStretching}
\end{equation}
Since $\log \rho_i$ follows a random process, the mean square increases linearly in time. The rescaled mean square $V/t$ reaches a saturation value equal to $V/t = 0.18 /\tau_\eta $. This is larger than the value $V/t = 0.1 /\tau_\eta $ found in Girimaji and Pope \cite{Girimaji1990}. This difference can be caused by the slightly different Reynolds numbers, differences in the spectral resolutions and the volume forcing methods ${\bm f}$ in Navier-Stokes equations (\ref{eq:nse}). Note, that we inject kinetic energy at a constant rate into the flow which determines $\langle\epsilon\rangle$ \cite{Schumacher2007}. This differs to the stochastic forcing used in ref. \cite{Girimaji1990}. The third central moment of $\log\rho_i$ rescaled by $V^{3/2}$ corresponds to the skewness of the PDF of $\log \rho_i$
\begin{equation}
    S = \frac{\langle(\log \rho_i - \lambda t )^3\rangle}{V^{3/2}}.
    \label{eq:Skewness}
\end{equation}
Figure~\ref{fig:pdf_rho_and_moments}(d), the skewness of $\log \rho$ is initially weakly positive and decreases with time. It is curious to see that the PDFs of the surface stretching are skewed whereas the PDFs of line stretching (for pair dispersion) are not skewed. This effect is visible in the results of Girimaji and Pope \cite{Girimaji1990}. It should be noted that the PDFs of surface stretching are not skewed for a sine flow \cite{Meunier2022}.
 
The Taylor-microscale Reynolds number $Re_{\lambda}$ of the turbulent flow is given by 
\begin{equation}
\mathrm{Re}_\lambda = \sqrt{\frac{5}{3\nu \langle \epsilon\rangle}}\, U^2\,,
\end{equation}
and has a moderate value of 52 as provided in table \ref{table:turb_cases}. The spectral resolution is $k_{\rm max}\eta_K\approx 6$ for $Sc=0.7$ and $k_{\rm max}\eta_B\approx 3$ for $Sc=70$ with the maximally resolved wavenumber $k_{\rm max}=2\pi\sqrt{2}N/(3L)$. This guarantees a sufficient resolution of the smallest scales of the turbulent flow. The DNS are performed for two Schmidt numbers of 0.7 and 70. 
\begin{table}[b!]
\begin{tabular}{cccccccccccccc}
\hline\hline
Run & $L\;$ [m]      & $N^3$    & $\Delta x/\eta_{K}$ & $\delta A_0\;$ [mm$^2$] & $\ell_0\;$ [m] &$Re$  & $Re_\lambda$& $Sc$ & $\tau_\eta$ [s] & D [m$^2$s$^{-1}$]\\
\hline
T1 & 0.256 & $512^3$  & 0.5   & $1\, (=\eta_K^2)$       & $0.005$ & 1634  & 52 & 0.7 & 0.07 & 2.52$\cdot$10$^{-5}$\\ 
T2 & 0.256 & $2048^3$ & 0.125 & $1\, (=\eta_K^2)$       & $0.005$ & 1634  & 52 & 70  & 0.07 & 2.52$\cdot$10$^{-6}$\\ 
\hline\hline
\end{tabular}
\caption{Parameters of the direct numerical simulations. Here, $N^3$ is the number of grid points, $\Delta x/\eta_{K}$ is the ratio of the uniform grid spacing to Kolmogorov length. The energy injection rate is $\epsilon_{\rm inj}\approx0.004\,\text{m}^2\text{s}^{-3}$, which results in Kolmogorov microscale $\eta_K=1\,\text{mm}$. The root mean square velocity is $U=0.097$ m/s. Thus the large-scale eddy turnover time follows to $\tau_L=L/U=2.64$ s for both runs. The number of diffuselets is $N_{\rm diff}=256^2$.}
\label{table:turb_cases}
\end{table}   
\begin{figure}
\centering
\includegraphics[width=0.8 \linewidth]{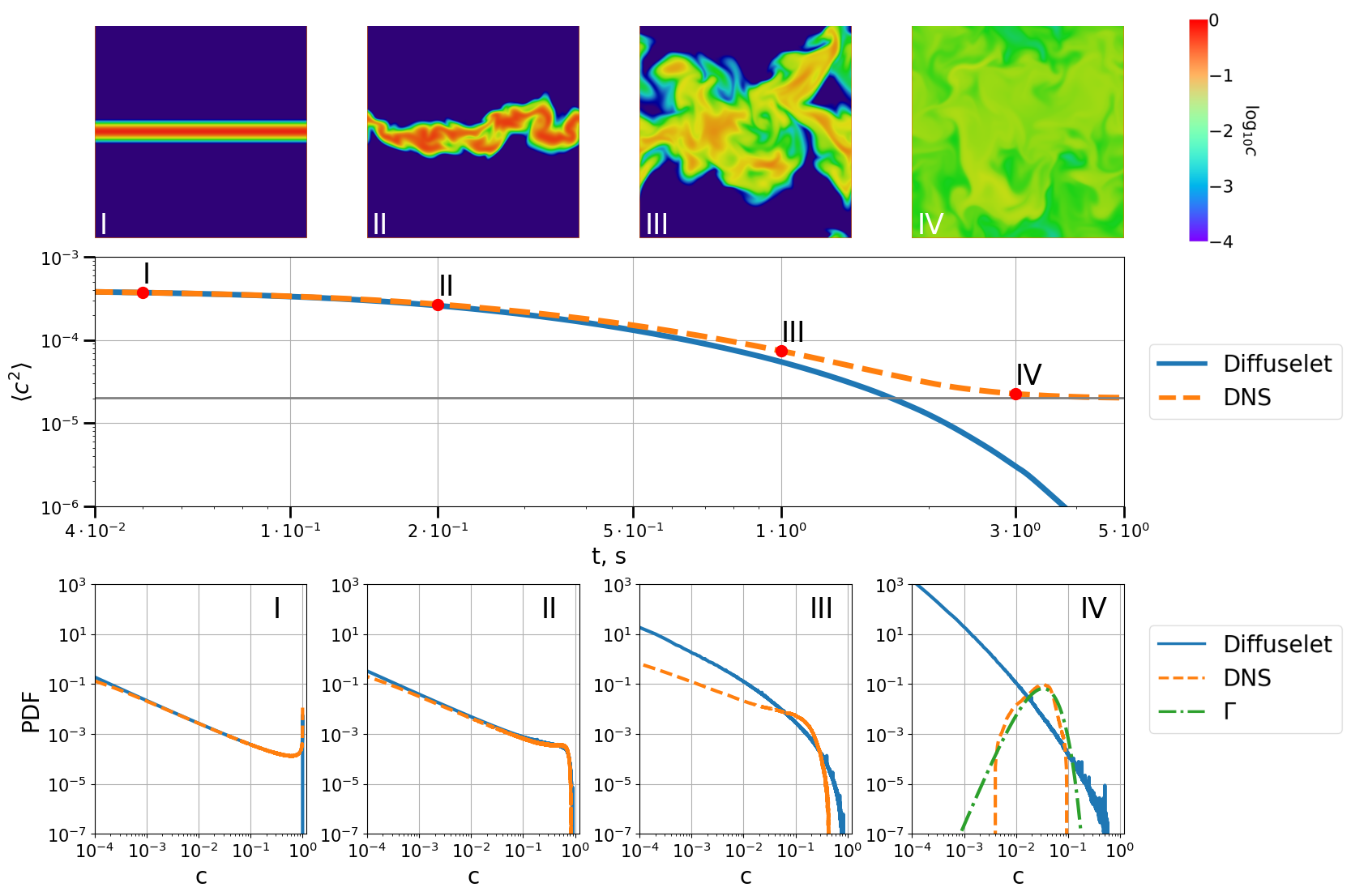}
\caption{Comparison of the diffuselet method with DNS for a turbulent flow at $Sc=0.7$  (Run T1). The central panel presents the evolution of mean squared scalar  $\langle c^2\rangle$ over time, with selected time points (I–IV) marked by red dots. The top row shows contour plots of the scalar field (logarithm of concentration) at these times, illustrating the effect of high diffusivity on scalar structures. The bottom row compares PDF $P(c)$ of scalar concentration $c$ for both methods, showing their agreement across different stages of scalar mixing. Green dashed-dotted line represents gamma distribution, see eq. \eqref{gamma_dis}, around the mean value.}
\label{fig:cloud_results}
\end{figure}

\subsubsection{Schmidt number $Sc=0.7$}    
The comparison of DNS run T1 with the diffuselet model is presented in Fig. \ref{fig:cloud_results}. The central part of figure displays again mean squared scalar  $\langle c^2(t)\rangle$ which corresponds to the supersaturation field in the cloud slab. The top row shows contour plots of the concentration field at selected times, illustrating the dynamics of the mixing process. The bottom row compares the PDFs of the passive scalar, obtained by both methods, at the same time instants. It is observed how the filament is increasingly deformed and convoluted while simultaneously being subject to molecular diffusion. Both, DNS and diffuselet methods initially show a very good agreement, as demonstrated by mean square and PDF. Deviations arise for $t\gtrsim 1$ s, which corresponds to times larger than $\sim 0.3 \tau_L$ and thus to instant III in the figure. The DNS data converge to the perfectly mixed state $\langle c^2\rangle\to \langle c\rangle^2$ for times $t\gg 1$ s. This is a consequence of the conserved scalar dynamics, which implies that the volume mean $\langle c\rangle = $ const. with respect to time $t$ in the present setup with periodic boundary conditions in all three spatial directions. Such an offset is absent in the diffuselet model (which is formulated without boundary conditions), such that the mean square continues here to decay exponentially with respect to time. The mixing time definition \eqref{eq:mixing_time} is adapted to the turbulent case where surfaces grow exponentially in time
\begin{equation}
    t_s = \frac{1}{2 \gamma_{\rm turb}}\log\left(\frac{\gamma_{\rm turb}\, \ell_0^2}{D}\right) \quad \mbox{with}\quad \gamma_{\rm turb}=\sqrt{\frac{\epsilon_{\rm in}}{2\nu}}=\sqrt{\frac{\langle\epsilon\rangle}{2\nu}} \,.
\label{eq:mixing_time1}
\end{equation}
Here, the stretching rate $\gamma_{\rm turb}$ is obtained for a statistically steady regime of the fluid turbulence with $\epsilon_{\rm in}=\langle \epsilon\rangle$. For run T1, this results to $\gamma_{\rm turb}=10.6$  s$^{-1}$ and thus a short mixing time of $t_s\approx 0.11$ s since $D=2.5 \cdot 10^{-5} \ \mathrm{m \ s}^{-2}$. The mixing time corresponds approximately to time instant II in Fig. \ref{fig:cloud_results}. It explains why at time instant II the mean square has started to decay. It also explains why the peak at $c=1$ has a disappeared in the concentration PDF. There is a very good agreement between the diffuselet and the DNS at this time instant.

However, there is a discrepancy at later times for the PDFs between the DNS and the diffuselet method due to periodic boundary conditions and the aggregation process. The scalar concentration PDF at time instant IV can be fitted fairly well to a gamma distribution (except for the far tails), which suggests that a scalar sheet aggregation process is at work, despite an absence of a viscous-convective cascade range for $Sc<1$. The gamma distribution has the following form
\begin{equation}
    P(c)=\frac{c^{\alpha-1}\theta^{\alpha}\exp(-\theta c)}{\Gamma(\alpha)}\,,
\label{gamma_dis}
\end{equation}
where $\alpha=\langle c\rangle^2/\langle c^2\rangle$ and $\theta=\langle c\rangle/\langle c^2\rangle=\alpha/\langle c\rangle$. By application of the substitution, $P(c) dc=P(\tilde c) d\tilde c$, we can convert the PDF in \eqref{gamma_dis} into
\begin{equation}
    P\left(\tilde{c}=\frac{c}{\langle c\rangle}\right)=\frac{\tilde{c}^{\alpha-1}\alpha^{\alpha}\exp(-\alpha \tilde{c})}{\Gamma(\alpha)}\,,
\label{gamma_dis1}
\end{equation}
with only one time-dependent fit parameter $\alpha$ left, which is interpreted as the average number of sheet aggregations \cite{Villermaux2003}. This is at time instant IV roughly just $\alpha\gtrsim 1$ for $Sc=0.7$.

Finally, we note that the convergence into the well-mixed final state can be postponed by a reduction of $\langle c\rangle$, i.e., by a further reduction of the initial filament sheet thickness $\ell_0$.

\begin{figure}
\centering
\includegraphics[width=0.84\linewidth]{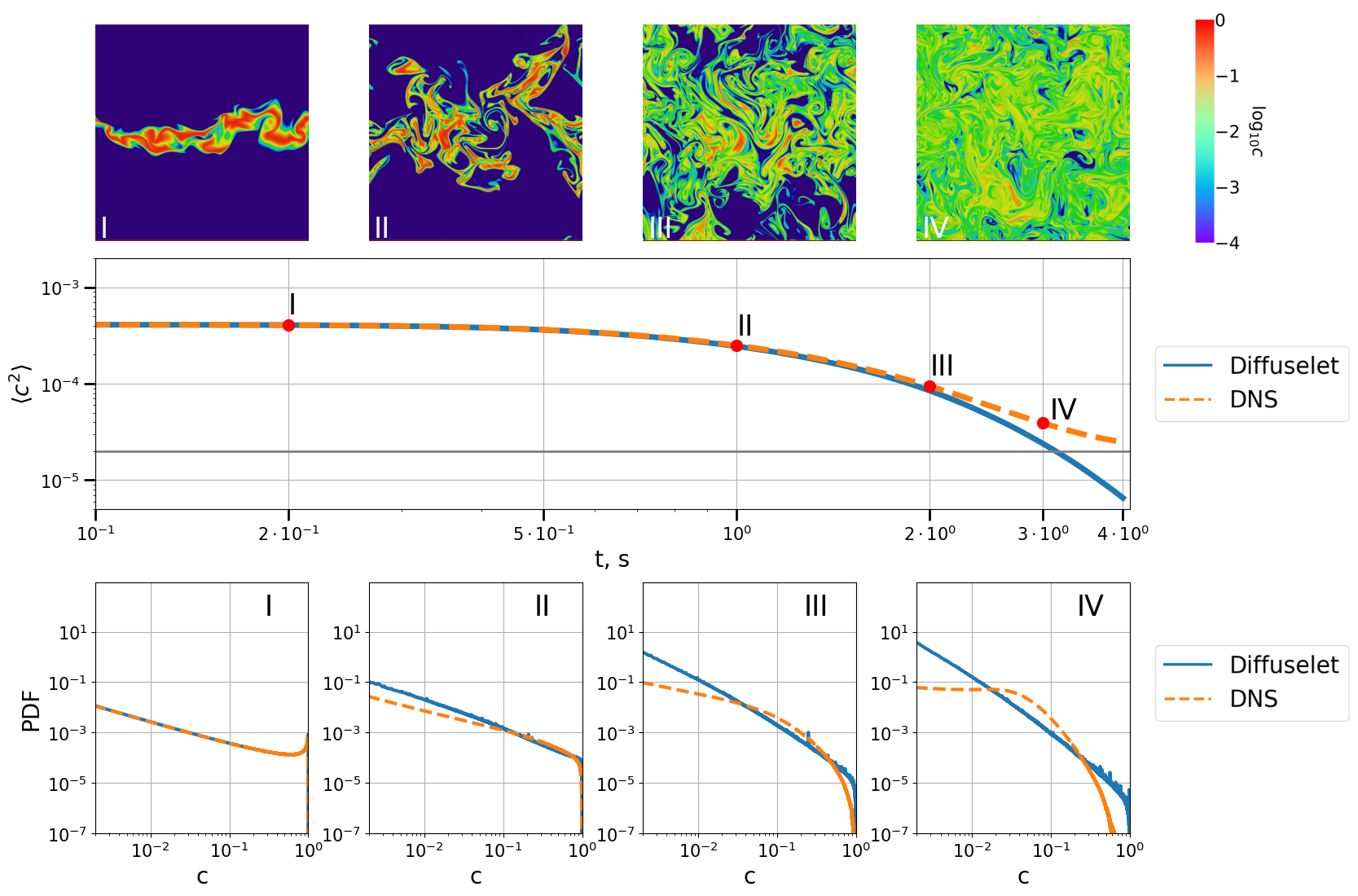}
\caption{Comparison of the diffuselet method with DNS for a turbulent flow at $Sc=70$  (Run T2). The central panel presents the evolution of mean squared scalar  $\langle c^2\rangle$ over time, with selected time points (I–IV) marked by red dots. The top row shows contour plots of the scalar field (logarithm of concentration) at these times, illustrating the effect of high diffusivity on scalar structures. The bottom row compares PDF of scalar concentration c for both methods, showing their agreement across different stages of scalar mixing.}
\label{fig:res_sc_70}
\end{figure}
    
\subsubsection{Schmidt number $Sc=70$}
In order to proceed with a comparison of the DNS and diffuselet method, we now move to a Schmidt number of $Sc=70$, which is two orders of magnitude larger than in subsection 3. It is clear that this case is not relevant for cloud turbulence. However, it allows us to further explore the capabilities of the diffuselet framework. The high Schmidt number causes a Batchelor length $\eta_B$ of the passive scalar, which is by approximately an order of magnitude smaller than $\eta_K$.  As shown in Fig. \ref{fig:res_sc_70}, structures of the scalar field are less blurred and show much finer striations in comparison to those of Fig. \ref{fig:cloud_results} at $Sc=0.7$. Consequently, the PDF and mean square of the scalar field require a longer time interval to converge to the well-mixed equilibrium state. As also evident from Fig. \ref{fig:res_sc_70}, both quantities agree for a longer time period. However, eventually also this DNS case starts to deviate from the predictions of the diffuselet model. 
    
The DNS for the present case were expensive since the full viscous-convective range has to be resolved for the high-Schmidt-number case. While leaving the outer box dimension $L$ at the same value as for $Sc=0.7$, the uniform mesh was by a factor of $4^3=64$ larger as seen in table \ref{table:turb_cases}. Therefore, with a further increase of the Schmidt number to values $Sc\sim 10^3$, the diffuselet method can serve as a potentially efficient way to model the turbulent mixing without the necessity to resolve all features of the scalar all the way down to the Batchelor scale. The model would then require information on the local flow stretching only, i.e., a resolution of the Kolmogorov length is sufficient. From this perspective, the diffuselet approach can be considered as an efficient Lagrangian subgrid-scale parametrization of the chaotic mixing in the viscous-convective scale range. Subgrid scale models for $Sc\gg 1$ are rare. We mention the nonlinear large-eddy simulation approach of Burton and Dahm \cite{Burton2005} that applies the multifractal model of kinetic energy and scalar dissipation rates to close the subgrid scales stresses. This closure model was extended in ref. \cite{Burton2008} to high Schmidt numbers. 
    
\begin{table}
\begin{tabular}{lccccc}
\hline\hline
Run& $L\;$ [m]     & $N^3$    &  $n_{d}\;$ [cm$^{-3}$]    & $K\;$ [m$^2$/s] & ]$r_0\;$ [$\mu$m]\\
\hline
D1 & 0.256 & $512^3$  & 100         & $1.2\cdot 10^{-11}$  & 20\\
D2 & 0.256 & $512^3$  & 500         & $1.2\cdot 10^{-11}$  & 20\\ 
D3 & 0.256 & $512^3$  & 1000        & $1.2\cdot 10^{-11}$  & 20\\
D4 & 0.512 & $1024^3$ & 100         & $1.2\cdot 10^{-11}$  & 20\\
D5 & 0.512 & $1024^3$ & 500         & $1.2\cdot 10^{-11}$  & 20\\
D6 & 0.512 & $1024^3$ & 1000        & $1.2\cdot 10^{-11}$  & 20\\
D7 & 0.256 & $512^3$  & 100         & $1.2\cdot 10^{-10}$  & 20\\
D8 & 0.256 & $512^3$  & 100         & $1.2\cdot 10^{-11}$  & 10\\
\hline\hline
\end{tabular}
\caption{Parameters of the simulations for turbulent scalar transport with phase changes. Here $n_d$ refers to the number density of liquid droplets, $K$ is the evaporation constant, $r_0$ is the initial droplet radius. Here, the initial thickness of cloud profile is always $\ell_0=0.01$ m. The number of diffuselets is $N_{\rm diff}=256^2$ for runs D1, D2, D3, D7 and D8 and $N_{\rm diff}=512^2$ for runs D4, D5, and D6.}
\label{table:dr_cases}
\end{table}
    
\subsection{Cloudy air filament mixing in three-dimensional homogeneous isotropic flow}
\subsubsection{Cloud water droplet growth model}
We extend the passive scalar mixing study to the case with phase changes in the following. As mentioned in the introduction, the interaction between cloud water droplets and small-scale turbulence plays a crucial role for the dynamics and lifetime of the (warm) turbulent cloud as a whole. To this end, we extend our physical model by cloud water droplets. Droplets are initially seeded randomly inside of the supersaturated region, where $s>0$. Recall that the liquid water mixing ratio $q_l$ is not  included as an Eulerian field, but as an ensemble of $N_t$ point-like Lagrangian tracers that carry a radius coordinate $r_j(t)$ with $j=1,\dots ,N_t$. The droplet evaporation and growth  adds a sink and source term to the advection-diffusion equation (\ref{eq:transoprt_c}), respectively \cite{Celani2005,Corrsin1951,Sardina2015,Fries2021}. This leads to
\begin{equation}
        \frac{\partial s}{\partial t}+ ({\bm u}\cdot{\bm \nabla})s = D_{s} {\bm \nabla}^2 s - A \frac{4 \pi \rho_l K}{V_{\mathrm{cell}}} \sum_{m=1}^{\hat{N}}  r_m(t) s({\bm X}_{m}(t),t),
        \label{eq:supersat_eq}
\end{equation}
where $A$ (measured in m$^3$/kg) is a parameter which depends on thermodynamic and physical parameters, such as saturation water vapor pressure and latent heat \cite{Pushenko2024}. Furthermore, $\rho_l$ is the liquid water density, $K$ the evaporation constant (measured in m$^2$/s),  $V_{\mathrm{cell}}=(\Delta x)^3$ the mesh cell volume, $r_j$ the radius of the $j$-th cloud water droplet with spatial coordinates ${\bm X}_{j}$. The summation is over $\hat{N}$ droplets inside $V_{\rm cell}$ at position ${\bm x}$. Furthermore, $D_s\approx D$. The constant $K$ carries the physical dimension of a diffusion constant; it is a function of the latent heat, the saturation pressure, the environmental temperature in the vicinity of the droplet, see ref. \cite{Rogers1989} or \cite{Kumar2013} for a detailed derivation. The radius changes of the $j$-th droplet depends on the fluctuations of supersaturation in the vicinity of the droplet. This leads to \cite{Rogers1989}
\begin{equation}
\frac{d r_j^2}{dt} = 2 K s({\bm X}_{j}(t),t) \quad \mbox{for}\quad j=1,...,N_t\,.
\label{eq:radius_change}
\end{equation}
The equation describes a diffusional growth in the Lagrangian frame. The initial size of the droplets in the filament for the present case is uniformly set to $r_j(0)= 
r_0=20 \mu$m (except run D8 with $r_0=10 \mu$m) . The droplets are thus too large to incorporate curvature and solutal effects in \eqref{eq:radius_change} on the one hand. On the other hand, they are still too small for collision-coalescence. For simplicity, they will thus be considered as Lagrangian tracers. The droplet radius evolves due to the evaporation process and serves as a source of water vapor. This source term is the condensation date $C_d$, the last term on the right-hand size of eq. (\ref{eq:supersat_eq}) which is given by
\begin{equation}
    C_d ({\bm x},t)= \frac{4 \pi \rho_l K}{V_{\mathrm{cell}}} \sum_{m=1}^{\hat{N}}  r_m(t) s({\bm X}_{m}(t),t)\,.
\end{equation}
The condensation rate enters as a Eulerian field in \eqref{eq:supersat_eq} and is obtained from the Lagrangian values along each discrete droplet trajectory ${\bm X}(t)$ by reshuffling to the 8 neighboring sites ${\bm x}$ of the uniform computational grid, loosely speaking by an inverse interpolation. In this way, we couple Eulerian and Lagrangian model parts. The Schmidt number is again $Sc\approx 0.7$. Table \ref{table:dr_cases} summarizes all cases that are studied in this subsection. Characteristic cloud water droplet number densities are $n\sim 10^2$ cm$^{-3}$; we will however enhance this density to artificially high values to investigate the impact of a dense spray-like configuration on the mixing process. Following ref. \cite{Kumar2012}, we will also enhance the growth diffusion constant $K$ by an order of magnitude to quantify the impact of a stronger droplet response to subsaturation. Together with size effects, this results in 8 DNS runs to compare DNS and diffuselet approach. 

The mono-disperse droplets are seeded randomly in the center of the cloudy air filament for time $t=0$, at places where the vapor mixing ratio $q_v>q_{vs}$, i.e., where the supersaturation $s>0$. Note also that the supersaturation field $s({\bm x},t)$ is rescaled in the same way to $c({\bm x},t)$ by eq. \eqref{eq:scalar_rescale} as for the droplet-free mixing case.
   
\begin{figure}
\centering
\includegraphics[width=0.84\linewidth]{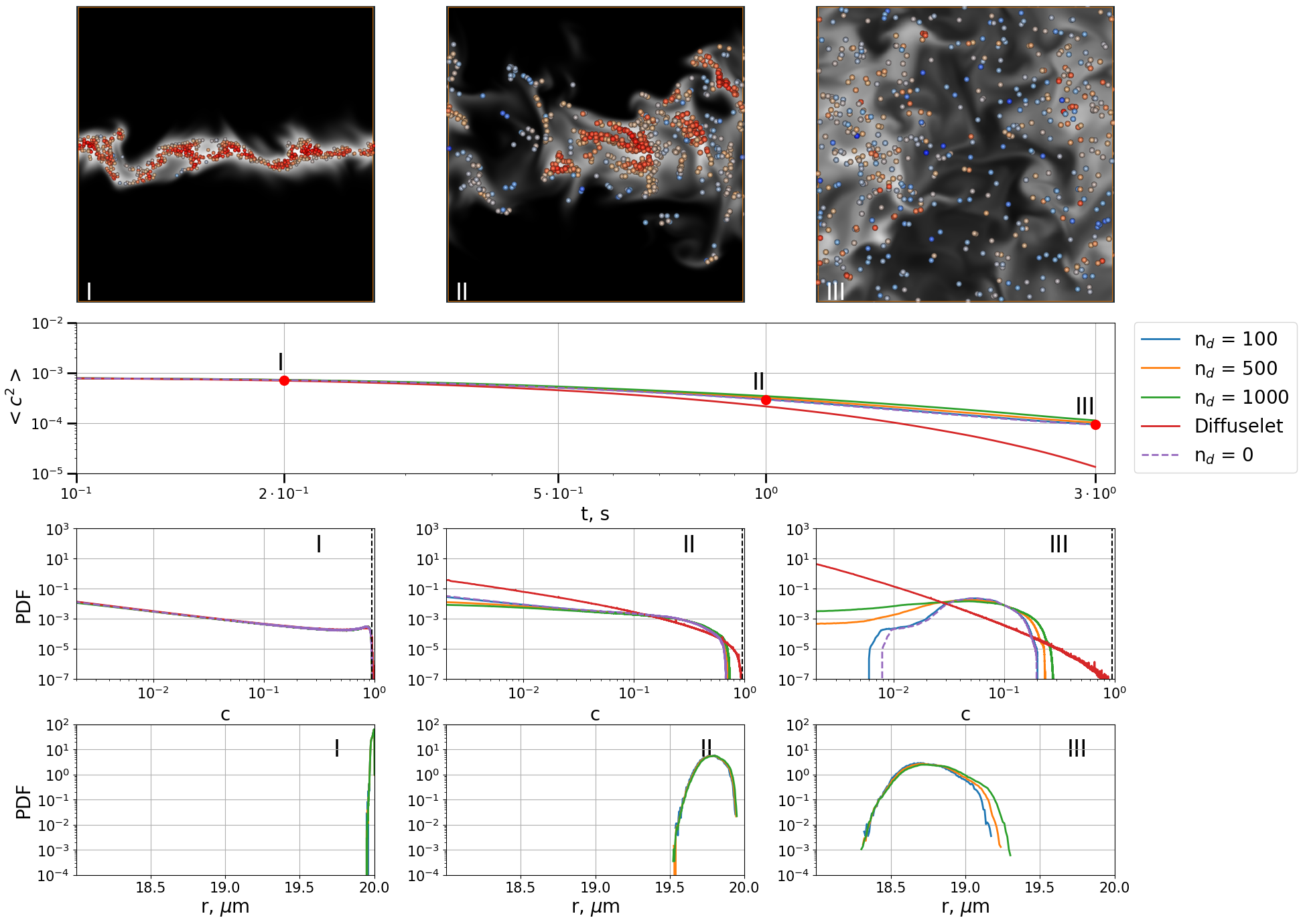}
\caption{Impact of additional cloud water droplets on the turbulent mixing process. First row: Vertical cuts through simulation domain snapshots at different time instants.  Droplets and scalar contours are shown for run D1 from table \ref{table:dr_cases}. The rescaled scalar supersaturation field of the filament and the seeded droplets are shown. The scalar is plotted in logarithmic contour levels. The droplets are colored with respect to their radius from red (largest) to blue (smallest). Second row: mean squared scalar  versus time, $\langle c^2(t)\rangle$. Third row:  Scalar PDFs at the three time instants from the top row. The dashed black line represents the critical value $c_c=1/(1+s_{max})$, which distinguishes regions where droplets grow from those where they shrink. Fourth row: Corresponding droplet size distributions. The number densities and the diffuselet case are indicated by the legend in the second row. These are runs D1, D2, and D3 from table \ref{table:dr_cases}.}
\label{fig:juicy_cases}
\end{figure}

\subsubsection{Comparison to diffuselet model for different cloud water droplet number densities}
The numerical simulations demonstrate that the additional source term due to cloud water droplets influences the statistical properties of the scalar field as shown in Fig. \ref{fig:juicy_cases}. For a number density $n_d=100$ cm$^{-3}$ in run D1, which is typical for a warm cloud, the difference to the previous droplet-free case in run T2 remains small for the whole time evolution that is considered. Deviations from the droplet-free case become larger for higher number densities in runs D2 and D3. More droplets are subject to evaporation which enhances the magnitude of the condensation rate; $C_d<0$ holds for evaporation which is a source of vapor content. Furthermore, the droplet size distribution is broadest for the highest number density, which is visible in the bottom row of Fig. \ref{fig:juicy_cases}. More droplets remain in the supersaturated fractions of the filaments such that their evaporation is delayed. The top row of the figure shows vertical cuts through the simulation volume displaying contours of the scalar field together with droplets in this cross section plane which are colored with respect to their individual size. Three time instants of $t=0.2$ s, 1 s, and 3 s are shown.  Since the droplets are Lagrangian tracers they do not decouple from the scalar filaments in this study. This is different to ref. \cite{Goetzfried2019} where gravitational settling and particle inertia were included. Blue colored satellite droplets are thus observable at the subsaturated outer scalar filaments that are subject to the strongest entrainment of clear subsaturated air. To conclude, the comparison with the diffuselet method is basically as for the droplet-free run T1 for the realistic value of the cloud droplet number density. Deviations increase with increasing $n_d$. 

To verify the influence of initial droplet size on the mixing process, an additional simulation with a reduced initial droplet radius of $r_0=10\,\mu\text{m}$  was performed (run D8 in table \ref{table:dr_cases}) for $n_d=100$ cm$^{-3}$. The mean square of the scalar and the PDF showed negligible differences between the two cases, indicating that scalar mixing remains largely unaffected by droplet size at this range. However, the droplet radius distributions increasingly deviate of each other with progressing time. A consequence of the droplet evaporation law \eqref{eq:radius_change} is that the initially smaller droplets shrink faster and thus lead to a broader size distribution at later times (not shown). This result is not shown, but is consistent with the trends discussed above.

\begin{figure}[t!]    
\centering
\includegraphics[width=0.8\linewidth]{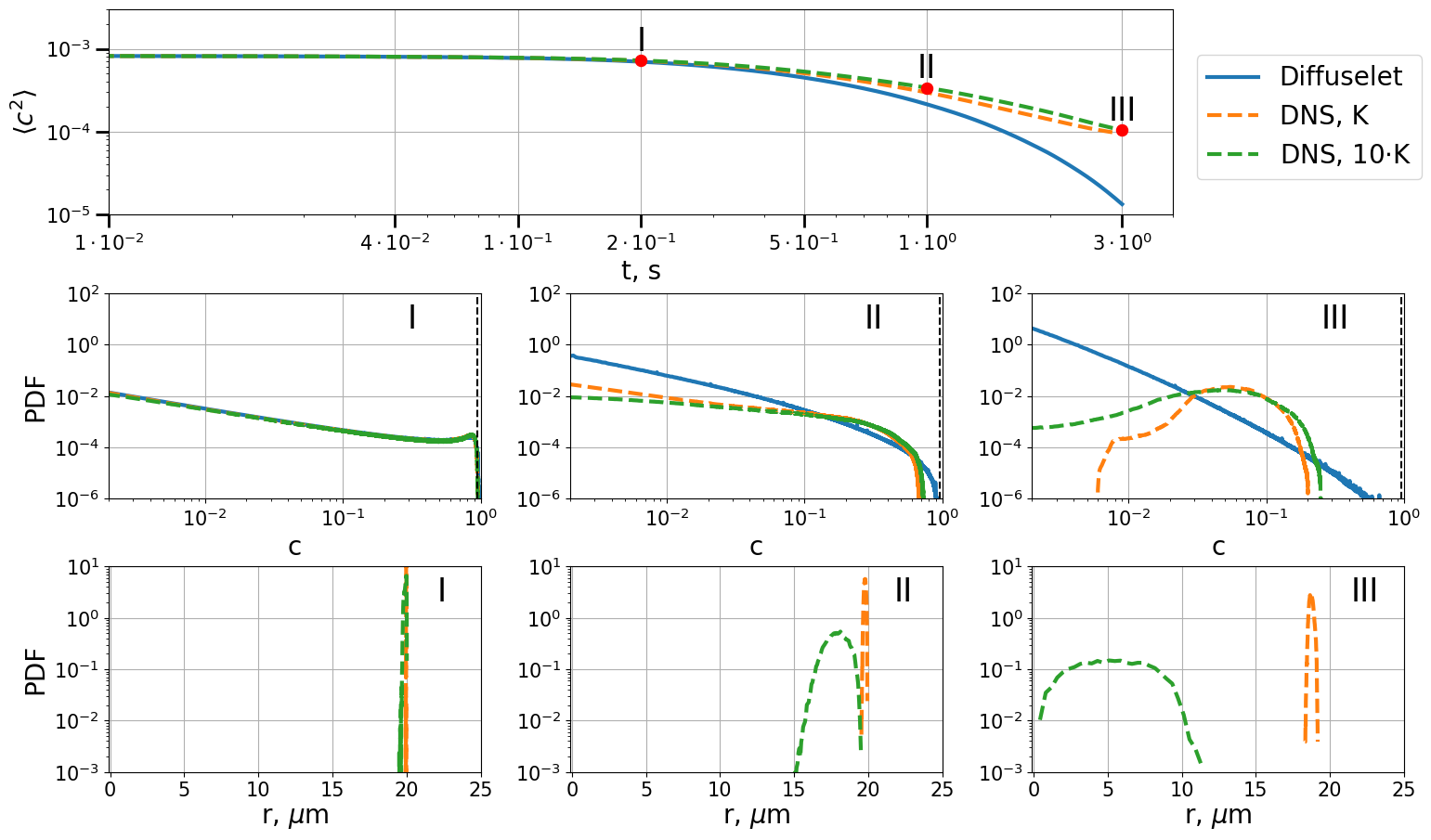}
\caption{Comparison of results between realistic (run D1) and artificially enhanced (run D7) constant $K$ in the droplet growth equation \eqref{eq:radius_change}. Top row: mean squared scalar  versus time, $\langle c^2(t)\rangle$. Second row:  Scalar PDFs at the three time instants from the top row. The dashed black line represents again the critical value $c_c=1/(1+s_{max})$, which distinguishes regions where droplets grow from those where they shrink. Bottom row: Corresponding droplet size distributions. The values of $K$ and the diffuselet case are indicated by the legend in the top row.}
\label{fig:enhanced_and_diff}
\end{figure}
    
\subsubsection{Comparison to diffuselet model for enhanced droplet evaporation}    
The main relative humidity (or saturation) for the specified configuration, even in a well-mixed equilibrium, is considerably less than $100 \%$, thereby  enabling droplets to evaporate further. Nevertheless, the rate of evaporation for realistic physical parameters does not permit the evaporation of more than $7\%$ of the initial size for the time interval that we considered. The evaporation  constant $K$ (which has the physical dimension of a diffusion constant)  determines the rate at which droplets exchange mass with the surrounding air, see again eq. \eqref{eq:radius_change}. 

To further enhance evaporation of cloud water droplets, $K$ can be increased beyond realistic values which was done for run D7 from table \ref{table:dr_cases}. An increase of $K$ by one order of magnitude is sufficient to produce notable changes in the evolution of droplet size distribution, as shown in ref. \cite{Kumar2012}. Figure \ref{fig:enhanced_and_diff} summarizes the comparison of runs D2 and D7. Starting with $t=2$ s, the number of droplets begins to decrease due to the complete droplet evaporation in D7. It can be observed in the bottom row of the figure that the PDF of the droplet size distribution evolves from an initial delta-type distribution to a broader one, similar to what was shown in \cite{Kumar2012}. Droplets outside the core of the cloudy-air slab shrink more rapidly. It is also seen that the deviations to the diffuselet model in D7 increase. The effect of the additional condensation rate term is again clearly visible in the scalar PDFs by a saturated tail at the smallest scalar concentrations, similar to the artificially enhanced droplet number densities in the third row of Fig. \ref{fig:juicy_cases}.  

\subsubsection{Effect of bigger simulation domain size on comparison to diffuselet model}
\begin{figure}[t!]
        \centering
        \includegraphics[width=0.84\linewidth]{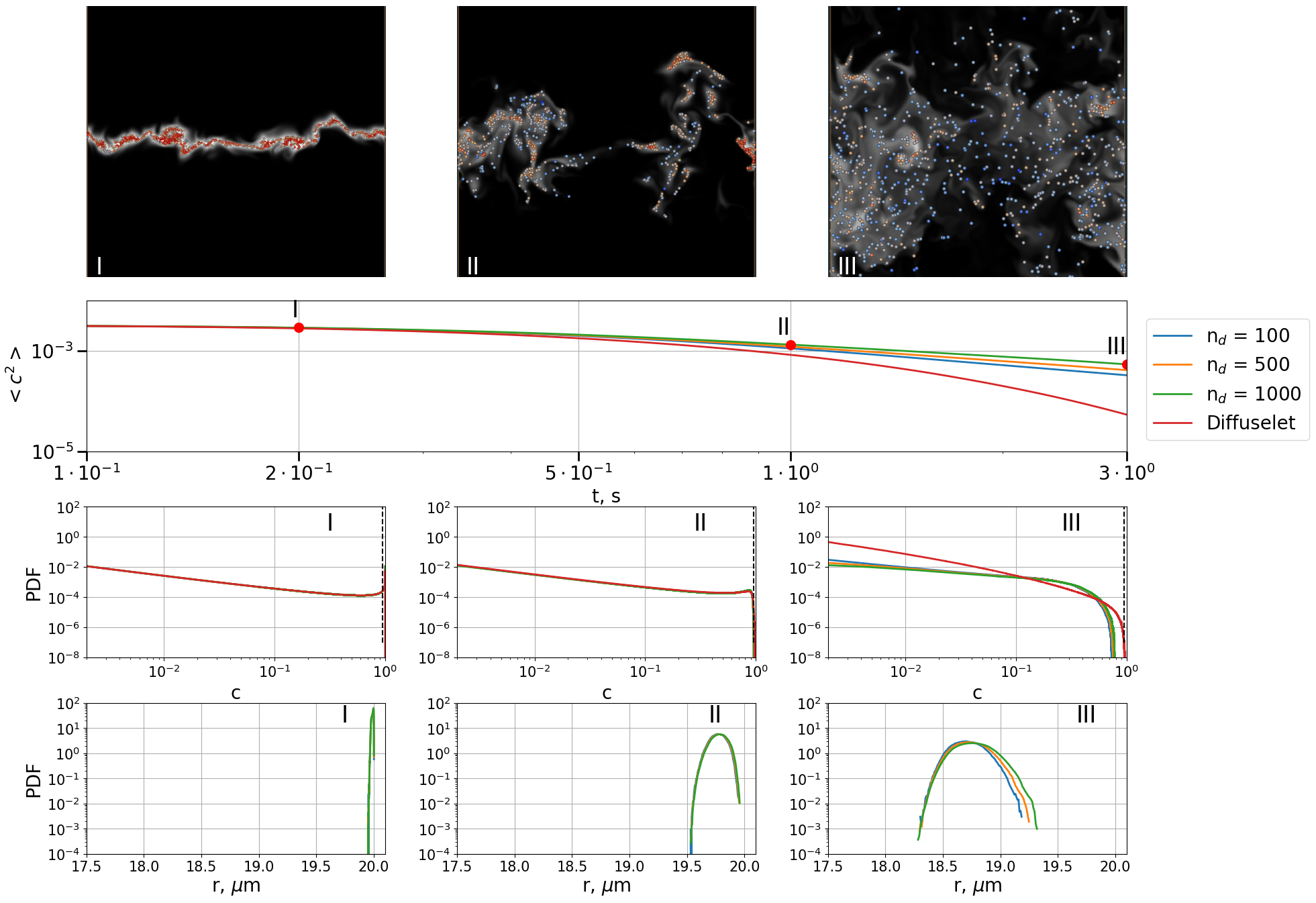}
        \caption{Same as in Fig. \ref{fig:juicy_cases}, but now for runs D4, D5, and D6 of table \ref{table:dr_cases} with box size $L=0.512$ m.}
        \label{fig:sprays_L_512}
    \end{figure}    
       
From the previous discussions, it becomes evident that for the given scalar diffusivity the ratio of filament to simulation volume (which is effectively $\ell_0/L$) is a factor that has to be included in the comparison of DNS and diffuselet methods. This ratio also determines the magnitude of the scalar mean value to which the mean squared scalar  relaxes in the final state of the turbulent mixing process. Runs D4 to D6 thus repeat runs D1 to D3, but in a simulation box $V=L^3$ that is twice as big in each space direction.   
Figure \ref{fig:sprays_L_512} confirms an improvement of the agreement of DNS results to those of the diffuselet method for this series when comparing to Fig. \ref{fig:juicy_cases}. 

\subsubsection{Time evolution for individual cloud water droplets}
The evaporation and size changes of the droplets can be approximated by the well-known d$^2$-law, which suggests that the square of the droplet diameter changes linearly with time. This result can be obtained from eq. (\ref{eq:radius_change}) as $d^2(t) \approx d_0^2 - 8 K t \langle s\rangle_V$ when the supersaturation $s({\bm X}_i(t))\to \langle s\rangle_V$, i.e., a constant. Here $d$ is droplet diameter, $d_0$ the initial value ($r=d/2$ in eq. (\ref{eq:radius_change})). The lifetime of a single droplet such a uniform field is
\begin{equation}
    t_1=\frac{d_0^2}{8K\langle s\rangle_V}\,,
\end{equation}
where $\langle s\rangle_V$ is the volume mean of the supersaturation field. This law assumes an idealized system, where evaporation is specified only by water vapor diffusion in a homogeneous background, given here by the volume average of the supersaturation field, $\langle s\rangle_V$; fluctuations of humidity are disregarded. The $d^2$-law is likely to be violated when a droplet experiences large fluctuations of temperature and water vapor content in the field in which it evaporates, like those resulting from turbulent mixing in clouds, as first anticipated by Srivastava \cite{Srivastava1989}.

\begin{figure}[t!]
\centering
\includegraphics[width=0.7\linewidth]{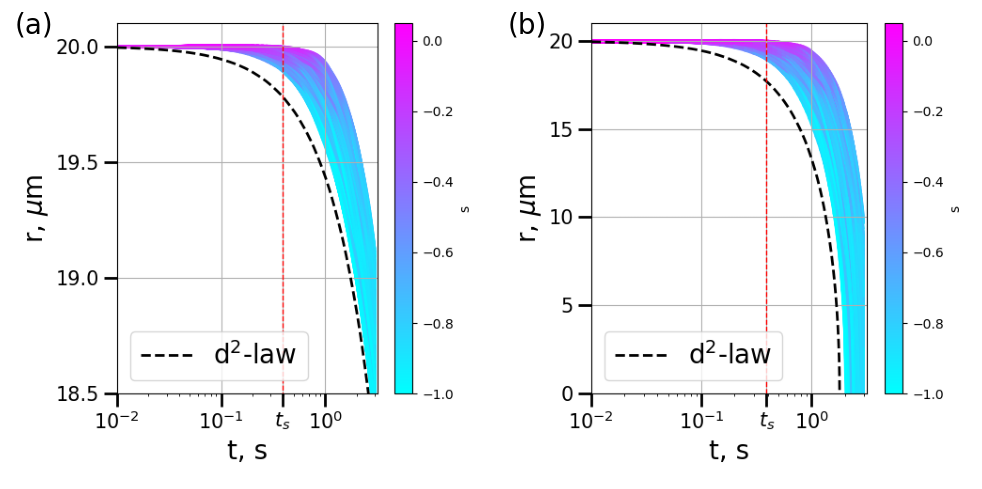}
\caption{Evaporation process along individual Lagrangian cloud water droplet tracks. The figure shows 1310 individual particles tracks colored with respect to time (see legend). (a) Run D1. (b) Run D7. See table \ref{table:dr_cases}. The black dashed line displays the  $d^2$-law for the evaporation in a homogeneous supersaturation fields. The red dashed line represents the mixing time $t_s$.}
\label{fig:d2-law}
\end{figure}

 Indeed, selected tracks of individual cloud water droplets $r(t)$ from the DNS data display a substantial evaporation delay as compared to the $d^2$ expectation, as can be seen in Fig. \ref{fig:d2-law}. The reason is easy to understand given our starting strip topology: close-packed droplets do not evaporate when surrounded by their own saturating vapor, but do so promptly (on a timescale given by the $d^2$-law) when their interstitial vapor concentration falls below saturation. The lifetime  of a whole droplets strip of width $\ell_0$ is thus given by the time $t_v$ it takes for the vapor to evacuate the droplets assembly, and thus coincides with the mixing time $t_s$ of the vapor constructed on the initial size of the strip \cite{Rivas2016,Villermaux2017} which reads
\begin{equation}
\begin{aligned}
    t_v&=\frac{1}{\gamma_{\rm turb}}\log\left(1+\phi^{-1}\right)\,,\\  
  \textrm{with}\quad\phi &= \frac{\rho_s-\rho_{\infty}}{\varphi\rho_l+(1-\varphi)\rho_s}\frac{2}{\sqrt{Pe}}\,,\\
     \textrm{and}\quad Pe&=\frac{\gamma_{\rm turb} \ell_0^2}{D},
    \label{eq:spray_time}
    \end{aligned}
\end{equation}
where $\rho_s$ is the saturated vapor density, $\rho_l$ the liquid water density, $\rho_{\infty}$ the vapor density outside of the spray, and $\varphi$ the liquid water volume fraction, $Pe=4208$ is the P\'eclet number. Recall that the stretching rate is found $\gamma_{\rm turb} = 10.6 \,\mathrm{s}^{-1}$, see directly below eq. \eqref{eq:mixing_time1}. The volume fraction $\varphi$ is defined as ratio of liquid water volume $V_l$ to vapor volume $V_v$ as
\begin{equation}
    \varphi = \frac{V_l}{V_v} = \frac{\sum_{i=1}^{N_t}4\pi r_i^3/3}{V_v},
    \label{eq:vol_frac}
\end{equation}
where $N_t$ is the total number of droplets. For case D1 of the present simulations, we have $\varphi\sim10^{-6}$, leading to $\phi\approx 4\cdot10^{-4}$, giving a spray lifetime calculated from eq. \eqref{eq:spray_time} of is $t_v = 0.73$ s in very fair agreement with the observations reported in Fig. \ref{fig:d2-law} showing that it takes roughly somewhat less than a second for most of the droplets to start to decrease rapidly. Evaporation of dense sprays is a mixing problem.

\begin{figure}[t!]
\centering
\includegraphics[width=0.9\linewidth]{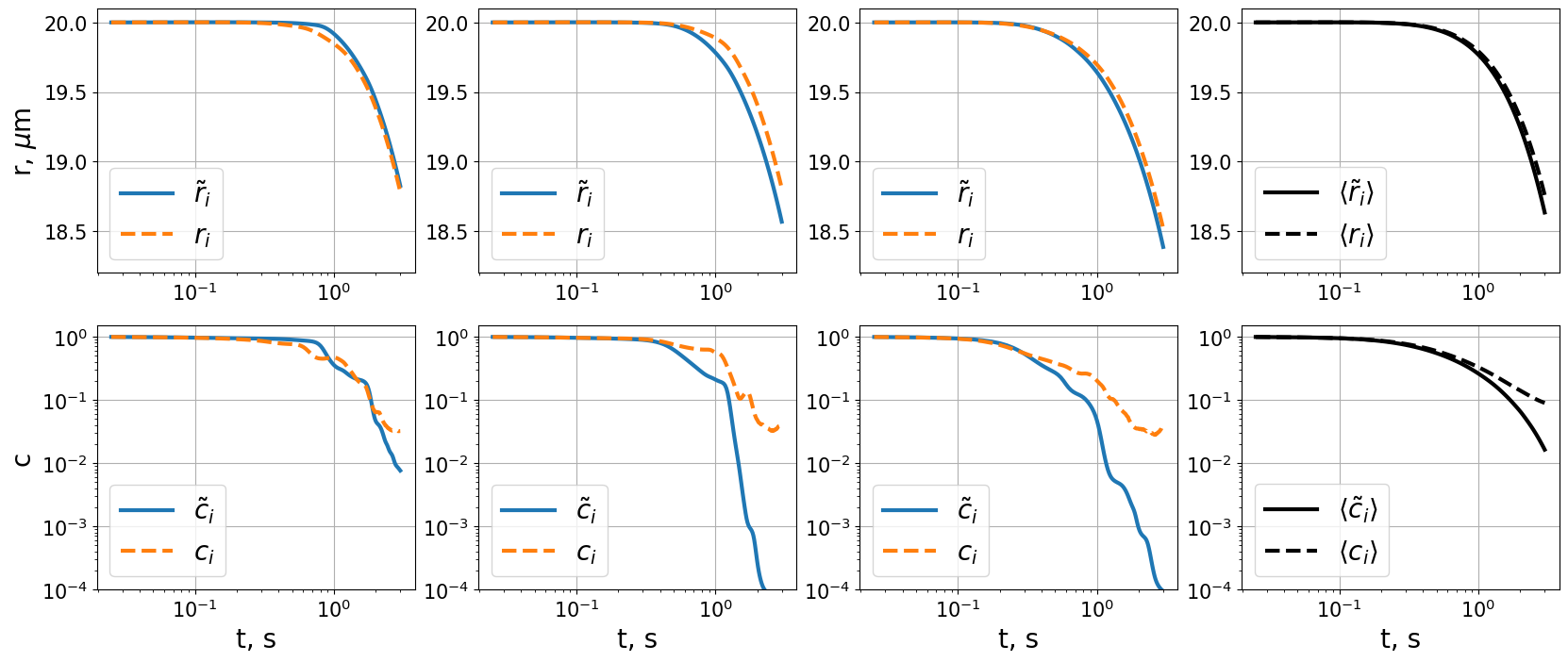}
\caption{Individual droplet evaporation histories directly obtained from the diffuselet model by eq. \eqref{eq:direct_evap} in comparison to the simulation results. This comparison is done for individual droplets. Three example cases are shown together with the average, that is taken over the whole droplet ensemble. Upper row: Time evolution of  droplet radius. Bottom row: evolution of scalar concentration. Quantities with a tilde correspond to results from the diffuselet model. Data are for run D1.}
\label{fig:diff_spray}
\end{figure}

\subsubsection{Evolution of individual cloud water droplets from the diffuselet model}
In subsection II B, we explained that the maximum concentration of the diffuselet $i$ is connected to the dimensionless time $\tau_i(t)$, see eq. \eqref{ranztime}, by the following relation 
\begin{equation}
\tilde{c}({\bm X}_i,t) \approx \frac{1}{\sqrt{\tau_i}}\,.
\label{eq:direct_evap}
\end{equation}
This suggests to investigate, how this direct evaluation of the supersaturation at the droplet position ${\bm X}_i(t)$ compares to the one which follows from the numerically simulated scalar field. We conduct this comparison for simulation run D1 at a Schmidt number of $Sc=0.7$ in the following. Thus, we determine $\tilde{s}({\bm X}_i,t)$ from $\tilde{c}({\bm X}_i,t)$ and inserting this into eq. \eqref{eq:radius_change} for each droplet. We note at this point, that the original diffuselet model does not include the phase changes which are studied here. Thus, we can expect deviations in the course of time.  
    
Figure \ref{fig:diff_spray} shows the comparison between both methods. As can be seen for all three examples, up to the mixing time $t_s$, both methods agree very well even for individual droplets. However, differences become visible at later times which either lead to a slightly faster or slightly slower decrease of the droplet radius with time. Similarly, the scalar concentration starts to differ, most probably caused by additional deformations of the diffuselet sheet in the two space directions normal to the cross-sheet coordinate. These effects are not fully captured by the model. When these data for individual droplets are averaged over the full droplet ensemble, we obtain again the results that have been reported already in Fig. \ref{fig:juicy_cases}. They are replotted here in the last column of Fig. \ref{fig:diff_spray}.  
    
\begin{figure}
\centering
\includegraphics[width=0.8\linewidth]{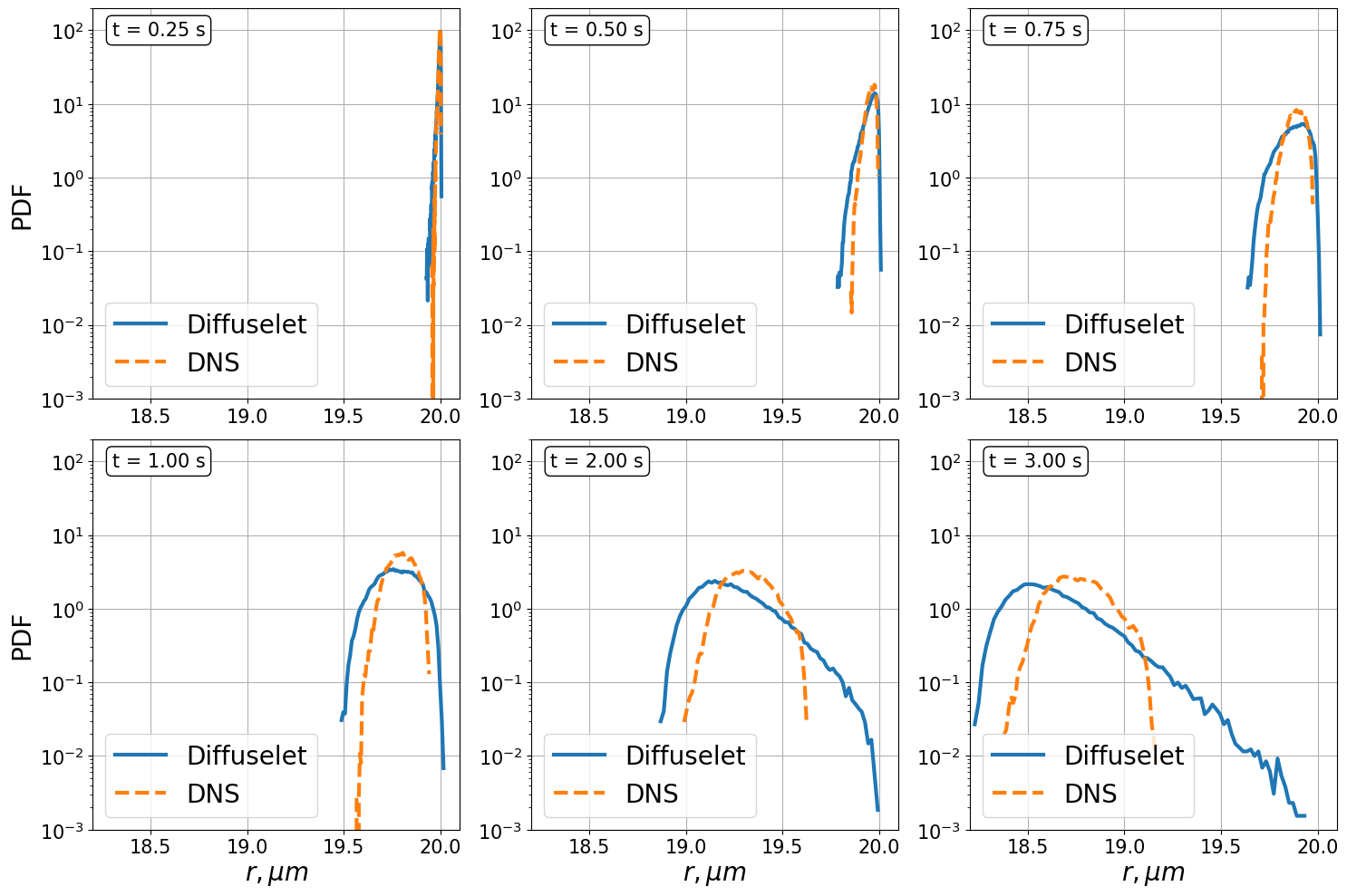}
\caption{Comparison of probability density function of the cloud water droplet radius obtained by diffuselet method (blue lines) and DNS (orange line) for six different time instants which are indicated in each panel. Data are for run D1.}
\label{fig:radius_pdfs}
\end{figure}

Finally, Fig. \ref{fig:radius_pdfs} shows the PDFs of the droplet radii at different times, comparing both, diffuselet method and DNS. At initial times for $t=0.25$ s and 0.5 s, the PDF agree fairly well. The distributions are still localized near initial radius $r_0 = 20 \; \mu$m. As time progresses,  for $t=0.75$ s and 1 s, differences between both methods start to occur, particularly in the left tail of the distribution for the droplets that evaporate fastest. For the last two instants at $t=2$ s and 3 s, a pronounced right tail is observed for the diffuselet method. The distribution obtained from the DNS model is narrower on both sides of the tails, which is a results of the saturation of the mean square to the well-mixed limit.      

\section{Conclusion}
The primary focus of this work was a comparison of the turbulent mixing of a scalar filament in a three-dimensional incompressible turbulent Navier-Stokes flow by two methods: the Lagrangian kinematic diffuselet model \cite{Meunier2022} and direct numerical simulations of scalar turbulence. The configuration of interest is a sheet-like thin scalar filament at the beginning of the mixing process. The turbulence is in statistically stationary regime. This filament (or sheet) is covered by little area elements, the diffuselets; their center is attached to a Lagrangian tracer which is advected in the turbulent flow. Along the individual diffuselet element trajectories, we then collect the individual stretching history which allows us to predict the time evolution of the mean squared scalar  and the probability density function PDF of the scalar concentration $c$ (and correspondingly $s$) of the mixing process in the control volume $V=L^3$. The results of the diffuselet framework and the DNS agree as long as the limitations of approximating the process in a box with periodic boundary conditions in the numerical simulations become important. The scalar mixing process is compared for two Schmidt numbers, $Sc=0.7$ and 70. We consider single-phase and two-phase turbulent mixing processes for this study. The latter case is a simple model for the entrainment and subsequent turbulent mixing of clear air with cloudy air (which is represented as the simple sheet-like filament) in a small control volume $V$ at the edge of a turbulent atmospheric cloud. In the two-phase case, cloud water droplets are seeded to the filament subject to slow shrinkage and eventual evaporation. The droplets are treated here simply as Lagrangian tracers; gravitational settling and droplet inertia are neglected in the present study. For a discussion of this simplification, we refer to \cite{Pushenko2024}.  

The major findings can be summarized as follows. The results of the diffuselet method and the DNS for the single-phase case agree well up to the mixing time scale $t_s$. Afterwards, deviations start to grow systematically. They are caused by the relaxation of the concentration fluctuations to the finite volumetric mean value of the passive scalar concentration field -- the well-mixed final state, which is not included in the diffuselet method. Here, the finite mean is a consequence of the conservation of concentration in the triply periodic control volume. The agreement improves when the Schmidt number is enhanced and when the initial filament thickness is smaller (not shown).  

The comparison of the diffuselet model to DNS for the cloudy air filament showed a level of agreement similar to that of a single-phase scalar field. This holds particularly for values of droplet number density and diffusion constant $K$ in the droplet growth equation that correspond to realistic values inside the cloud. The scalar field in this case is the supersaturation field $s$ which determines the local growth conditions of the cloud water droplets. For enhanced number density and droplet response to subsaturation, the deviations become larger since the magnitude of the source (or sink) term -- the condensation rate $C_d$ -- in the scalar transport eq. \eqref{eq:supersat_eq} grows. Again, we can conclude that a smaller initial cloudy air filament thickness relative to the box size improves the agreement to the diffuselet model. Turbulent fluctuations of the supersaturation field due to entrainment cause also deviations from the $d^2$-law for droplet evaporation in sprays. Our results thus agree with a recent mixing experiment in a laboratory flow \cite{Rivas2016}. To summarize this part, we demonstrated a first extension of the diffuselet method to an evaporating phase of a collection of closely packed droplets and anticipated the overall the overall evaporation time of a strip of a relatively dense spray, see eqns. \eqref{eq:spray_time}.

To conclude, we have thus demonstrated that the diffuselet method is applicable even for the case of low Schmidt numbers $Sc\sim 1$ including phase changes. Our work shows clearly that the diffuselet method has its limitations, in particular when longer evolution times $t>t_s$ have to be considered. By construction, the agreement to DNS is best for high-Schmidt-number mixing processes. Here, the model can be even used as a Lagrangian subgrid scale approach that models the mixing in the viscous-convective scale range. A further extension of the diffuselet method would consist in a direct incorporation of the phase changes, i.e., the integration of loss of liquid water for evaporation or the loss of vapor content for condensation into the kinematic Lagrangian framework. These processes would then introduce an active scalar component into the model with latent heat release and evaporative cooling. These efforts can be expected to be reported elsewhere. 

\acknowledgments
The work of V.P. and S.S. was supported by ITN CoPerMix. The project has received funding from the European Union’s Horizon 2020 research and innovation program under the Marie Skłodowska-Curie grant agreement N°956457. Supercomputing time has been provided at the University Computer Center (UniRZ) of the TU Ilmenau. The authors also gratefully acknowledge the Gauss Centre for Supercomputing e.V. (https://www.gauss-centre.eu) for funding this project by providing computing time on the GCS Supercomputer SuperMUC-NG at Leibniz Supercomputing Centre (https://www.lrz.de).

\begin{appendix} 
\section{Passive scalar mixing in steady analytical flow}
Here, we validate the diffuselet method for a simple analytical steady velocity field in a three-dimensional cubic domain with periodic boundary conditions
\begin{align}
u_x({\bm x},t) = A_1 \cos(\omega_1 y),\quad
u_y({\bm x},t) = A_2 \sin(\omega_2 x),\quad
u_z({\bm x},t) = A_3\,,
\label{eq:an_flow}
\end{align}
where $A_1$, $A_2$, $A_3$, as well as $\omega_1$, $\omega_2$ are constants. For simplicity, we consider $A_1=A_2=A=0.
1  \ \mathrm{m \ s}^{-1}$, $A_3=0$, together with $\omega_1 = \omega_2 = \omega = 2 \pi/L$. For convenience, we will also provide scales and velocities often in physical units, such that the reader gets a better impression about the dimensions of the considered mixing volumes. The diffusivity coefficient $D$ was chosen to be five times smaller than that of water vapor, so $D=5\cdot10^{-6} \ \mathrm{m \ s}^{-2}$. As mentioned above, the initial scalar field takes the form of a sheet with a transversal Gaussian profile and an initial thickness $\ell_0 = 0.005 \;\mathrm{m}$. This is about $2\%$ of the box size $L=0.512$ m. The sheet is divided into $512 \times 512$ surface elements with surface $\delta A_0$ and sheet thickness $\ell_0$; these are the diffuselets (see again Fig. \ref{fig:init_profile}). This implies an advection of 262,144 Lagrangian tracers simultaneously with the analytical flow. 
    
\begin{figure}
\centering
\includegraphics[width=0.8\linewidth]{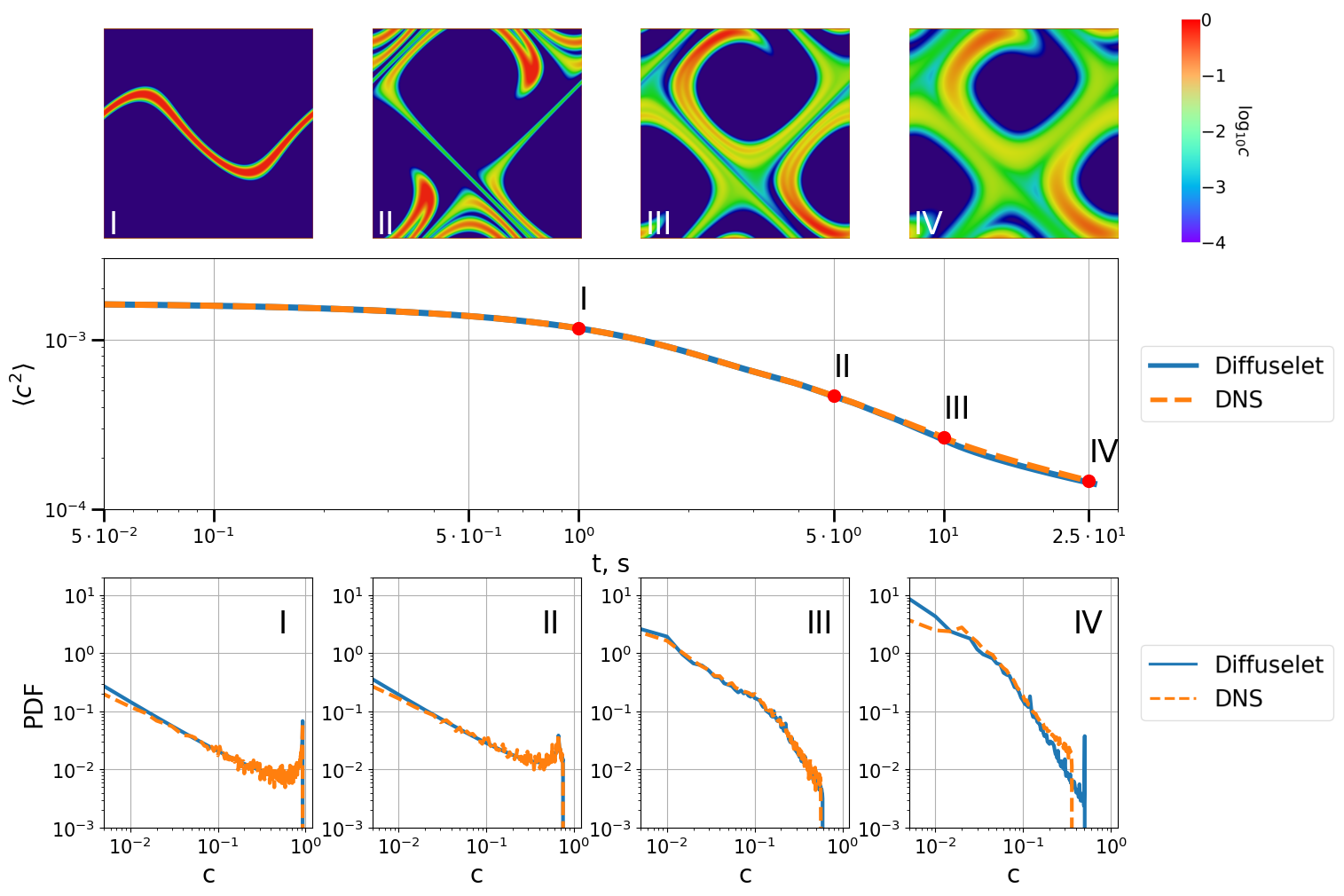}
\caption{Comparison of the diffuselet method with DNS for scalar transport in a steady analytical flow given by eq. \eqref{eq:an_flow}. The central panel shows the time evolution of the mean squared scalar $\langle c^2(t)\rangle$ in a doubly-logarithmic plot. The red markers indicate selected time instants (I–IV). The top row shows two-dimensional contour plots of the scalar field (logarithm of concentration) at these times. The bottom row of panels provides the corresponding PDFs of scalar concentration $c$ for both methods at the four times.}
\label{fig:an_flow_512}
\end{figure}

The comparison of the results for mean square and PDF of the scalar field $c$, obtained by the diffuselet method and the DNS, is presented in Fig. \ref{fig:an_flow_512}. The mean squared scalar  in the DNS, which is calculated as volume average at time $t$, coincides perfectly with the one from the diffuselet. The mean square  starts to decay after $t_s \approx 1$ s. At this mixing time, scalar diffusion is enhanced because the thickness of the sheet becomes smaller than the Batchelor scale. Here, the flow is 2D and steady, which leads to a linear growth in time of the sheet's surface as in a pure shear flow. The mixing time is thus given by \cite{Villermaux2019} 
\begin{equation}
    t_s = \frac{1}{\gamma} (3 Pe )^{1/3}    
\label{eq:mixing_time}
\end{equation}
where the P\'eclet number $Pe =\gamma \ell_0^2 / D$ is based on the initial thickness $\ell_0$ of the scalar sheet. Here, the shear $\gamma=2\pi A_1/L\approx 1.2 \;\mathrm{s}^{-1}$, see again eqs. \eqref{eq:an_flow}, which results in a mixing time $t_s\approx 2$ s, in good agreement with the numerical results. 
 
Fig. \ref{fig:an_flow_512} displays the concentration PDFs, calculated using the DNS and using the diffuselet model with Eq.~\eqref{eq:pdf_final}. There is an excellent agreement between the two methods up to times $t\lesssim 10$ s. Initially, the PDF exhibits a peak at $c=1$, corresponding to the maximum concentration inside the sheet of scalar. After the mixing time, this peak disappears and the PDFs become monotonic decreasing functions of $c$. After $t_{\mathrm{II}}=5$ s, the maximal concentration starts to decrease as a function of time, meaning that all parts of the sheet have started to diffuse. The maximum concentration is equal to $c_{\mathrm{max}}=0.5$ at $t_{\mathrm{III}}=5$ s. At a later time ($t_{\mathrm{IV}}=10$ s), the maximal concentration predicted by the diffuselet model is slightly larger than in the DNS. Indeed, this maximal concentration corresponds to weakly stretched scalar sheet for which the diffuselet method is inappropriate, since it models the diffusion as a 1D transverse diffusive process whereas these points diffuse in two directions in the DNS.

\section{Tests of numerical resolution}
\begin{figure}[h!]
\centering
\includegraphics[width=0.8\linewidth]{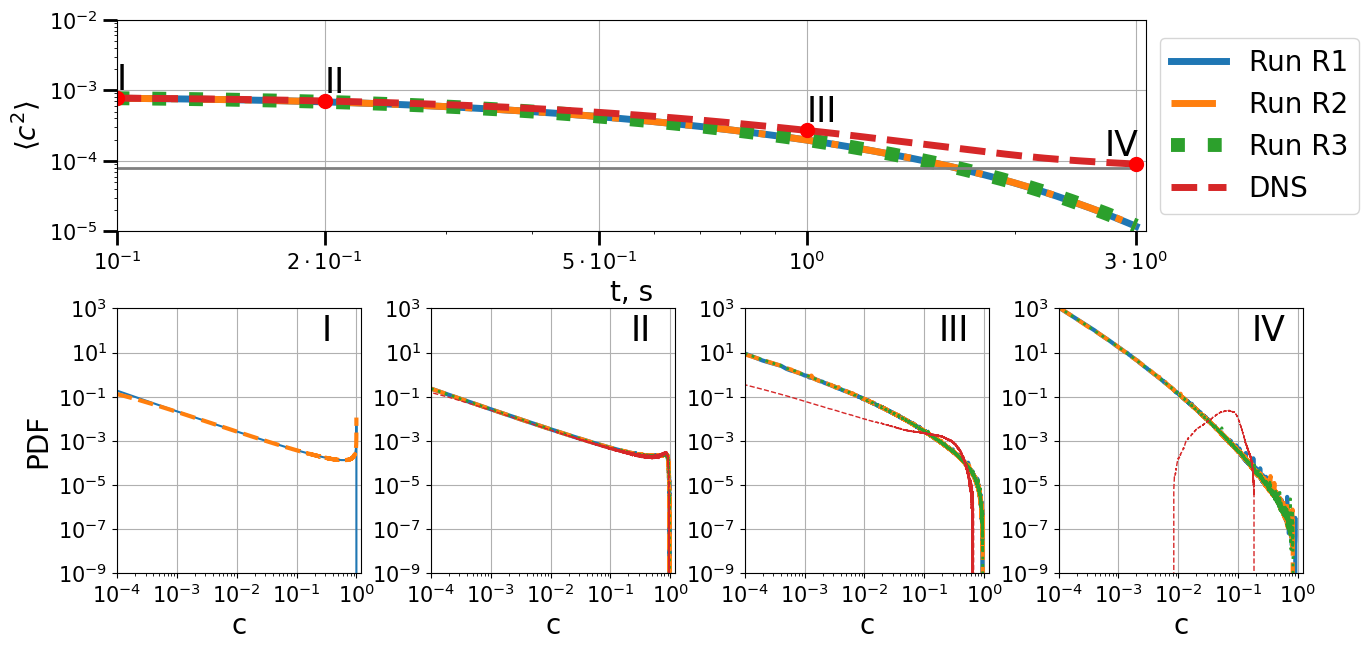}
\caption{Comparing results of the diffuselet method with differently coarse decompositions of the initial filament sheet. Top row: mean squared scalar  versus time. Bottom row: PDFs of the scalar concentration for 4 different instants. The initial diffuselet surface for runs R1 to R3 is indidcated in table \ref{table:comp_res}).}
\label{fig:comp_res}
\end{figure}

The  diffuselet method might be sensitive to the number of diffuselets that cover the filament sheet. In this appendix, we present additional runs (see table \ref{table:comp_res}) to investigate the performance for differently fine diffuselet decompositions. To this end, we compare three DNS runs R1, R2, and R3 with different spectral resolution and correspondingly different diffuselet numbers, all advecting the scalar in the same turbulent flow. We can conclude that the different decompositions lead to the same statistics, as seen in Fig. \ref{fig:comp_res}, for both, mean squared scalar  and scalar distribution.  

\begin{table}[t!]
\begin{tabular}{lcccccccc}
\hline\hline
Run&  $L\;$ [m]     & $N^3$    & $\Delta x/\eta_K$ &  $\delta A_0$ &  $\ell_0\;$ [m] & $Re$  &  $Re_\lambda$ & $Sc$\\
\hline
R1 & 0.256 & $256^3$  & 1.0   &$\Delta x\times  \Delta x$     & 0.01  & 1634 & 52 & 0.7 \\
R2 & 0.256 & $512^3$  & 0.5   &$2\Delta x\times 2\Delta x$  & 0.01  & 1634 & 52 & 0.7 \\ 
R3 & 0.256 & $1024^3$ & 0.25  &$4\Delta x\times 4\Delta x$ & 0.01  & 1634 & 52 & 0.7 \\
\hline\hline
\end{tabular}
\caption{Parameters of three DNS runs to compare differently fine decompositions of the initial filament into diffuselets together with differently fine numerical grids. The energy injection rate is $\epsilon_{\rm inj}\approx0.004\,\text{m}^2\text{s}^{-3}$, which results in Kolmogorov microscale $\eta_K=1\,\text{mm}$.}
\label{table:comp_res}
\end{table}
\end{appendix}

\bibliography{main_rev}

\providecommand{\noopsort}[1]{}\providecommand{\singleletter}[1]{#1}%
\begin{thebibliography}{48}%
\makeatletter
\providecommand \@ifxundefined [1]{%
 \@ifx{#1\undefined}
}%
\providecommand \@ifnum [1]{%
 \ifnum #1\expandafter \@firstoftwo
 \else \expandafter \@secondoftwo
 \fi
}%
\providecommand \@ifx [1]{%
 \ifx #1\expandafter \@firstoftwo
 \else \expandafter \@secondoftwo
 \fi
}%
\providecommand \natexlab [1]{#1}%
\providecommand \enquote  [1]{``#1''}%
\providecommand \bibnamefont  [1]{#1}%
\providecommand \bibfnamefont [1]{#1}%
\providecommand \citenamefont [1]{#1}%
\providecommand \href@noop [0]{\@secondoftwo}%
\providecommand \href [0]{\begingroup \@sanitize@url \@href}%
\providecommand \@href[1]{\@@startlink{#1}\@@href}%
\providecommand \@@href[1]{\endgroup#1\@@endlink}%
\providecommand \@sanitize@url [0]{\catcode `\\12\catcode `\$12\catcode `\&12\catcode `\#12\catcode `\^12\catcode `\_12\catcode `\%12\relax}%
\providecommand \@@startlink[1]{}%
\providecommand \@@endlink[0]{}%
\providecommand \url  [0]{\begingroup\@sanitize@url \@url }%
\providecommand \@url [1]{\endgroup\@href {#1}{\urlprefix }}%
\providecommand \urlprefix  [0]{URL }%
\providecommand \Eprint [0]{\href }%
\providecommand \doibase [0]{https://doi.org/}%
\providecommand \selectlanguage [0]{\@gobble}%
\providecommand \bibinfo  [0]{\@secondoftwo}%
\providecommand \bibfield  [0]{\@secondoftwo}%
\providecommand \translation [1]{[#1]}%
\providecommand \BibitemOpen [0]{}%
\providecommand \bibitemStop [0]{}%
\providecommand \bibitemNoStop [0]{.\EOS\space}%
\providecommand \EOS [0]{\spacefactor3000\relax}%
\providecommand \BibitemShut  [1]{\csname bibitem#1\endcsname}%
\let\auto@bib@innerbib\@empty
\bibitem [{\citenamefont {Shaw}(2003)}]{Shaw2003}%
  \BibitemOpen
  \bibfield  {author} {\bibinfo {author} {\bibfnamefont {R.~A.}\ \bibnamefont {Shaw}},\ }\bibfield  {title} {\bibinfo {title} {Particle-turbulence interactions in atmospheric clouds},\ }\href {https://doi.org/10.1146/annurev.fluid.35.101101.161125} {\bibfield  {journal} {\bibinfo  {journal} {Annu. Rev. Fluid Mech.}\ }\textbf {\bibinfo {volume} {35}},\ \bibinfo {pages} {183} (\bibinfo {year} {2003})}\BibitemShut {NoStop}%
\bibitem [{\citenamefont {Bodenschatz}\ \emph {et~al.}(2010)\citenamefont {Bodenschatz}, \citenamefont {Malinowski}, \citenamefont {Shaw},\ and\ \citenamefont {Stratmann}}]{Bodenschatz2010}%
  \BibitemOpen
  \bibfield  {author} {\bibinfo {author} {\bibfnamefont {E.}~\bibnamefont {Bodenschatz}}, \bibinfo {author} {\bibfnamefont {S.~P.}\ \bibnamefont {Malinowski}}, \bibinfo {author} {\bibfnamefont {R.~A.}\ \bibnamefont {Shaw}},\ and\ \bibinfo {author} {\bibfnamefont {F.}~\bibnamefont {Stratmann}},\ }\bibfield  {title} {\bibinfo {title} {Can we understand clouds without turbulence?},\ }\href {https://doi.org/10.1126/science.1185138} {\bibfield  {journal} {\bibinfo  {journal} {Science}\ }\textbf {\bibinfo {volume} {327}},\ \bibinfo {pages} {970} (\bibinfo {year} {2010})}\BibitemShut {NoStop}%
\bibitem [{\citenamefont {Stevens}\ and\ \citenamefont {Bony}(2013)}]{Stevens2013}%
  \BibitemOpen
  \bibfield  {author} {\bibinfo {author} {\bibfnamefont {B.}~\bibnamefont {Stevens}}\ and\ \bibinfo {author} {\bibfnamefont {S.}~\bibnamefont {Bony}},\ }\bibfield  {title} {\bibinfo {title} {Water in the atmosphere},\ }\href {https://doi.org/10.1063/PT.3.2009} {\bibfield  {journal} {\bibinfo  {journal} {Phys. Today}\ }\textbf {\bibinfo {volume} {66}},\ \bibinfo {pages} {29} (\bibinfo {year} {2013})}\BibitemShut {NoStop}%
\bibitem [{\citenamefont {Vogel}\ \emph {et~al.}(2022)\citenamefont {Vogel}, \citenamefont {Albright}, \citenamefont {Vial}, \citenamefont {George}, \citenamefont {Stevens},\ and\ \citenamefont {Bony}}]{Vogel2022}%
  \BibitemOpen
  \bibfield  {author} {\bibinfo {author} {\bibfnamefont {R.}~\bibnamefont {Vogel}}, \bibinfo {author} {\bibfnamefont {A.~L.}\ \bibnamefont {Albright}}, \bibinfo {author} {\bibfnamefont {J.}~\bibnamefont {Vial}}, \bibinfo {author} {\bibfnamefont {G.}~\bibnamefont {George}}, \bibinfo {author} {\bibfnamefont {B.}~\bibnamefont {Stevens}},\ and\ \bibinfo {author} {\bibfnamefont {S.}~\bibnamefont {Bony}},\ }\bibfield  {title} {\bibinfo {title} {Strong cloud–circulation coupling explains weak trade cumulus feedback},\ }\href {https://doi.org/10.1038/s41586-022-05364-y} {\bibfield  {journal} {\bibinfo  {journal} {Nature}\ }\textbf {\bibinfo {volume} {612}},\ \bibinfo {pages} {696} (\bibinfo {year} {2022})}\BibitemShut {NoStop}%
\bibitem [{\citenamefont {Rogers}\ and\ \citenamefont {Yau}(1989)}]{Rogers1989}%
  \BibitemOpen
  \bibfield  {author} {\bibinfo {author} {\bibfnamefont {R.~R.}\ \bibnamefont {Rogers}}\ and\ \bibinfo {author} {\bibfnamefont {M.~K.}\ \bibnamefont {Yau}},\ }\href@noop {} {\emph {\bibinfo {title} {A short course in cloud physics}}}\ (\bibinfo  {publisher} {Butterworth-Heinemann, Oxford, UK},\ \bibinfo {year} {1989})\BibitemShut {NoStop}%
\bibitem [{\citenamefont {Baker}\ \emph {et~al.}(1984)\citenamefont {Baker}, \citenamefont {Breidenthal}, \citenamefont {Choularton},\ and\ \citenamefont {Latham}}]{Baker1984}%
  \BibitemOpen
  \bibfield  {author} {\bibinfo {author} {\bibfnamefont {M.~B.}\ \bibnamefont {Baker}}, \bibinfo {author} {\bibfnamefont {R.~E.}\ \bibnamefont {Breidenthal}}, \bibinfo {author} {\bibfnamefont {T.~W.}\ \bibnamefont {Choularton}},\ and\ \bibinfo {author} {\bibfnamefont {J.}~\bibnamefont {Latham}},\ }\bibfield  {title} {\bibinfo {title} {The effects of turbulent mixing in clouds},\ }\href {https://doi.org/10.1175/1520-0469(1984)041%3C0299:TEOTMI%3E2.0.CO;2} {\bibfield  {journal} {\bibinfo  {journal} {J. Atmos. Sci.}\ }\textbf {\bibinfo {volume} {41}},\ \bibinfo {pages} {299} (\bibinfo {year} {1984})}\BibitemShut {NoStop}%
\bibitem [{\citenamefont {Andrejczuk}\ \emph {et~al.}(2006)\citenamefont {Andrejczuk}, \citenamefont {Grabowski}, \citenamefont {Malinowski},\ and\ \citenamefont {Smolarkiewicz}}]{Andrejczuk2006}%
  \BibitemOpen
  \bibfield  {author} {\bibinfo {author} {\bibfnamefont {M.}~\bibnamefont {Andrejczuk}}, \bibinfo {author} {\bibfnamefont {W.~W.}\ \bibnamefont {Grabowski}}, \bibinfo {author} {\bibfnamefont {S.~P.}\ \bibnamefont {Malinowski}},\ and\ \bibinfo {author} {\bibfnamefont {P.~K.}\ \bibnamefont {Smolarkiewicz}},\ }\bibfield  {title} {\bibinfo {title} {{Numerical simulation of cloud--clear air interfacial mixing: Effects on cloud microphysics}},\ }\href {https://doi.org/10.1175/JAS3813.1} {\bibfield  {journal} {\bibinfo  {journal} {J. Atmos. Sci.}\ }\textbf {\bibinfo {volume} {63}},\ \bibinfo {pages} {3204} (\bibinfo {year} {2006})}\BibitemShut {NoStop}%
\bibitem [{\citenamefont {Celani}\ \emph {et~al.}(2008)\citenamefont {Celani}, \citenamefont {Mazzino},\ and\ \citenamefont {Tizzi}}]{Celani2008}%
  \BibitemOpen
  \bibfield  {author} {\bibinfo {author} {\bibfnamefont {A.}~\bibnamefont {Celani}}, \bibinfo {author} {\bibfnamefont {A.}~\bibnamefont {Mazzino}},\ and\ \bibinfo {author} {\bibfnamefont {M.}~\bibnamefont {Tizzi}},\ }\bibfield  {title} {\bibinfo {title} {The equivalent size of cloud condensation nuclei},\ }\href {https://doi.org/10.1088/1367-2630/10/7/075021} {\bibfield  {journal} {\bibinfo  {journal} {New J. Phys.}\ }\textbf {\bibinfo {volume} {10}},\ \bibinfo {pages} {075021} (\bibinfo {year} {2008})}\BibitemShut {NoStop}%
\bibitem [{\citenamefont {Sardina}\ \emph {et~al.}(2015)\citenamefont {Sardina}, \citenamefont {Picano}, \citenamefont {Brandt},\ and\ \citenamefont {Caballero}}]{Sardina2015}%
  \BibitemOpen
  \bibfield  {author} {\bibinfo {author} {\bibfnamefont {G.}~\bibnamefont {Sardina}}, \bibinfo {author} {\bibfnamefont {F.}~\bibnamefont {Picano}}, \bibinfo {author} {\bibfnamefont {L.}~\bibnamefont {Brandt}},\ and\ \bibinfo {author} {\bibfnamefont {R.}~\bibnamefont {Caballero}},\ }\bibfield  {title} {\bibinfo {title} {Continuous growth of droplet size variance due to condensation in turbulent clouds},\ }\href {https://doi.org/10.1103/PhysRevLett.115.184501} {\bibfield  {journal} {\bibinfo  {journal} {Phys. Rev. Lett.}\ }\textbf {\bibinfo {volume} {115}},\ \bibinfo {pages} {{184501}} (\bibinfo {year} {2015})}\BibitemShut {NoStop}%
\bibitem [{\citenamefont {Kumar}\ \emph {et~al.}(2018)\citenamefont {Kumar}, \citenamefont {G{\"o}tzfried}, \citenamefont {Suresh}, \citenamefont {Schumacher},\ and\ \citenamefont {Shaw}}]{Kumar2018}%
  \BibitemOpen
  \bibfield  {author} {\bibinfo {author} {\bibfnamefont {B.}~\bibnamefont {Kumar}}, \bibinfo {author} {\bibfnamefont {P.}~\bibnamefont {G{\"o}tzfried}}, \bibinfo {author} {\bibfnamefont {N.}~\bibnamefont {Suresh}}, \bibinfo {author} {\bibfnamefont {J.}~\bibnamefont {Schumacher}},\ and\ \bibinfo {author} {\bibfnamefont {R.~A.}\ \bibnamefont {Shaw}},\ }\bibfield  {title} {\bibinfo {title} {Scale dependence of cloud microphysical response to turbulent entrainment and mixing},\ }\href {https://doi.org/10.1029/2018MS001487} {\bibfield  {journal} {\bibinfo  {journal} {J. Adv. Model. Earth Syst.}\ }\textbf {\bibinfo {volume} {10}},\ \bibinfo {pages} {2777} (\bibinfo {year} {2018})}\BibitemShut {NoStop}%
\bibitem [{\citenamefont {Fries}\ \emph {et~al.}(2021)\citenamefont {Fries}, \citenamefont {Sardina}, \citenamefont {Svensson},\ and\ \citenamefont {Mehlig}}]{Fries2021}%
  \BibitemOpen
  \bibfield  {author} {\bibinfo {author} {\bibfnamefont {J.}~\bibnamefont {Fries}}, \bibinfo {author} {\bibfnamefont {G.}~\bibnamefont {Sardina}}, \bibinfo {author} {\bibfnamefont {G.}~\bibnamefont {Svensson}},\ and\ \bibinfo {author} {\bibfnamefont {B.}~\bibnamefont {Mehlig}},\ }\bibfield  {title} {\bibinfo {title} {Key parameters for droplet evaporation and mixing at the cloud edge},\ }\href {https://doi.org/10.1002/qj.4015} {\bibfield  {journal} {\bibinfo  {journal} {Q. J. R. Meteorol. Soc.}\ }\textbf {\bibinfo {volume} {147}},\ \bibinfo {pages} {2160} (\bibinfo {year} {2021})}\BibitemShut {NoStop}%
\bibitem [{\citenamefont {Kumar}\ \emph {et~al.}(2021)\citenamefont {Kumar}, \citenamefont {Ranjan}, \citenamefont {Yau}, \citenamefont {Bera},\ and\ \citenamefont {Rao}}]{Kumar2021}%
  \BibitemOpen
  \bibfield  {author} {\bibinfo {author} {\bibfnamefont {B.}~\bibnamefont {Kumar}}, \bibinfo {author} {\bibfnamefont {R.}~\bibnamefont {Ranjan}}, \bibinfo {author} {\bibfnamefont {M.~K.}\ \bibnamefont {Yau}}, \bibinfo {author} {\bibfnamefont {S.}~\bibnamefont {Bera}},\ and\ \bibinfo {author} {\bibfnamefont {S.~A.}\ \bibnamefont {Rao}},\ }\bibfield  {title} {\bibinfo {title} {Impact of high and low vorticity turbulence on cloud environment mixing and cloud microphysics processes},\ }\href {https://doi.org/10.5194/acp-21-12317-2021} {\bibfield  {journal} {\bibinfo  {journal} {Atmos. Chem. Phys.}\ }\textbf {\bibinfo {volume} {21}},\ \bibinfo {pages} {12317} (\bibinfo {year} {2021})}\BibitemShut {NoStop}%
\bibitem [{\citenamefont {Grabowski}\ \emph {et~al.}(2022)\citenamefont {Grabowski}, \citenamefont {Thomas},\ and\ \citenamefont {Kumar}}]{Grabowski2022}%
  \BibitemOpen
  \bibfield  {author} {\bibinfo {author} {\bibfnamefont {W.~W.}\ \bibnamefont {Grabowski}}, \bibinfo {author} {\bibfnamefont {L.}~\bibnamefont {Thomas}},\ and\ \bibinfo {author} {\bibfnamefont {B.}~\bibnamefont {Kumar}},\ }\bibfield  {title} {\bibinfo {title} {{Impact of cloud-base turbulence on CCN activation: single-size CCN}},\ }\href {https://doi.org/10.1175/JAS-D-21-0184.1} {\bibfield  {journal} {\bibinfo  {journal} {J. Atmos. Sci.}\ }\textbf {\bibinfo {volume} {79}},\ \bibinfo {pages} {551} (\bibinfo {year} {2022})}\BibitemShut {NoStop}%
\bibitem [{\citenamefont {Fries}\ \emph {et~al.}(2023)\citenamefont {Fries}, \citenamefont {Sardina}, \citenamefont {Svensson}, \citenamefont {Pumir},\ and\ \citenamefont {Mehlig}}]{Fries2023}%
  \BibitemOpen
  \bibfield  {author} {\bibinfo {author} {\bibfnamefont {J.}~\bibnamefont {Fries}}, \bibinfo {author} {\bibfnamefont {G.}~\bibnamefont {Sardina}}, \bibinfo {author} {\bibfnamefont {G.}~\bibnamefont {Svensson}}, \bibinfo {author} {\bibfnamefont {A.}~\bibnamefont {Pumir}},\ and\ \bibinfo {author} {\bibfnamefont {B.}~\bibnamefont {Mehlig}},\ }\bibfield  {title} {\bibinfo {title} {Lagrangian supersaturation fluctuations at the cloud edge},\ }\href {https://doi.org/10.1103/PhysRevLett.131.254201} {\bibfield  {journal} {\bibinfo  {journal} {Phys. Rev. Lett.}\ }\textbf {\bibinfo {volume} {131}},\ \bibinfo {pages} {254201} (\bibinfo {year} {2023})}\BibitemShut {NoStop}%
\bibitem [{\citenamefont {Magaritz-Ronen}\ \emph {et~al.}(2016)\citenamefont {Magaritz-Ronen}, \citenamefont {Pinsky},\ and\ \citenamefont {Khain}}]{Magaritz2016}%
  \BibitemOpen
  \bibfield  {author} {\bibinfo {author} {\bibfnamefont {L.}~\bibnamefont {Magaritz-Ronen}}, \bibinfo {author} {\bibfnamefont {M.}~\bibnamefont {Pinsky}},\ and\ \bibinfo {author} {\bibfnamefont {A.}~\bibnamefont {Khain}},\ }\bibfield  {title} {\bibinfo {title} {Drizzle formation in stratocumulus clouds: Effects of turbulent mixing},\ }\href {https://doi.org/10.5194/acp-16-1849-2016} {\bibfield  {journal} {\bibinfo  {journal} {Atmos. Chem. Phys.}\ }\textbf {\bibinfo {volume} {16}},\ \bibinfo {pages} {1849} (\bibinfo {year} {2016})}\BibitemShut {NoStop}%
\bibitem [{\citenamefont {Dimotakis}(2005)}]{Dimotakis2005}%
  \BibitemOpen
  \bibfield  {author} {\bibinfo {author} {\bibfnamefont {P.~E.}\ \bibnamefont {Dimotakis}},\ }\bibfield  {title} {\bibinfo {title} {Turbulent mixing},\ }\href {https://doi.org/10.1146/annurev.fluid.36.050802.122015} {\bibfield  {journal} {\bibinfo  {journal} {Annu. Rev. Fluid Mech.}\ }\textbf {\bibinfo {volume} {37}},\ \bibinfo {pages} {329} (\bibinfo {year} {2005})}\BibitemShut {NoStop}%
\bibitem [{\citenamefont {Sreenivasan}\ and\ \citenamefont {Schumacher}(2010)}]{Sreenivasan2010}%
  \BibitemOpen
  \bibfield  {author} {\bibinfo {author} {\bibfnamefont {K.~R.}\ \bibnamefont {Sreenivasan}}\ and\ \bibinfo {author} {\bibfnamefont {J.}~\bibnamefont {Schumacher}},\ }\bibfield  {title} {\bibinfo {title} {Lagrangian views on turbulent mixing of passive scalars},\ }\href {https://doi.org/10.1098/rsta.2009.0140} {\bibfield  {journal} {\bibinfo  {journal} {Phil. Trans. R. Soc. A}\ }\textbf {\bibinfo {volume} {368}},\ \bibinfo {pages} {1561} (\bibinfo {year} {2010})}\BibitemShut {NoStop}%
\bibitem [{\citenamefont {Sreenivasan}(2019)}]{Sreenivasan2019}%
  \BibitemOpen
  \bibfield  {author} {\bibinfo {author} {\bibfnamefont {K.~R.}\ \bibnamefont {Sreenivasan}},\ }\bibfield  {title} {\bibinfo {title} {{Turbulent mixing: A perspective}},\ }\href {https://doi.org/10.1073/pnas.1800463115} {\bibfield  {journal} {\bibinfo  {journal} {Proc. Natl. Acad. Sci. USA}\ }\textbf {\bibinfo {volume} {116}},\ \bibinfo {pages} {18175} (\bibinfo {year} {2019})}\BibitemShut {NoStop}%
\bibitem [{\citenamefont {Villermaux}(2019)}]{Villermaux2019}%
  \BibitemOpen
  \bibfield  {author} {\bibinfo {author} {\bibfnamefont {E.}~\bibnamefont {Villermaux}},\ }\bibfield  {title} {\bibinfo {title} {Mixing versus stirring},\ }\href {https://doi.org/10.1146/annurev-fluid-010518-040306} {\bibfield  {journal} {\bibinfo  {journal} {Annu. Rev. Fluid Mech.}\ }\textbf {\bibinfo {volume} {51}},\ \bibinfo {pages} {245} (\bibinfo {year} {2019})}\BibitemShut {NoStop}%
\bibitem [{\citenamefont {Batchelor}(1959)}]{Batchelor1959}%
  \BibitemOpen
  \bibfield  {author} {\bibinfo {author} {\bibfnamefont {G.~K.}\ \bibnamefont {Batchelor}},\ }\bibfield  {title} {\bibinfo {title} {{Small-scale variation of convected quantities like temperature in turbulent fluid Part 1. General discussion and the case of small conductivity}},\ }\href {https://doi.org/10.1017/S002211205900009X} {\bibfield  {journal} {\bibinfo  {journal} {J. Fluid Mech.}\ }\textbf {\bibinfo {volume} {5}},\ \bibinfo {pages} {113} (\bibinfo {year} {1959})}\BibitemShut {NoStop}%
\bibitem [{\citenamefont {Kraichnan}(1968)}]{Kraichnan1968}%
  \BibitemOpen
  \bibfield  {author} {\bibinfo {author} {\bibfnamefont {R.~H.}\ \bibnamefont {Kraichnan}},\ }\bibfield  {title} {\bibinfo {title} {Small-scale structure of a scalar field convected by turbulence},\ }\href {https://doi.org/10.1063/1.1692063} {\bibfield  {journal} {\bibinfo  {journal} {Phys. Fluids}\ }\textbf {\bibinfo {volume} {11}},\ \bibinfo {pages} {945} (\bibinfo {year} {1968})}\BibitemShut {NoStop}%
\bibitem [{\citenamefont {Villermaux}\ and\ \citenamefont {Duplat}(2003)}]{Villermaux2003}%
  \BibitemOpen
  \bibfield  {author} {\bibinfo {author} {\bibfnamefont {E.}~\bibnamefont {Villermaux}}\ and\ \bibinfo {author} {\bibfnamefont {J.}~\bibnamefont {Duplat}},\ }\bibfield  {title} {\bibinfo {title} {Mixing as an aggregation process},\ }\href {https://doi.org/10.1103/PhysRevLett.91.184501} {\bibfield  {journal} {\bibinfo  {journal} {Phys. Rev. Lett.}\ }\textbf {\bibinfo {volume} {91}},\ \bibinfo {pages} {184501} (\bibinfo {year} {2003})}\BibitemShut {NoStop}%
\bibitem [{\citenamefont {Meunier}\ and\ \citenamefont {Villermaux}(2003)}]{Meunier2003}%
  \BibitemOpen
  \bibfield  {author} {\bibinfo {author} {\bibfnamefont {P.}~\bibnamefont {Meunier}}\ and\ \bibinfo {author} {\bibfnamefont {E.}~\bibnamefont {Villermaux}},\ }\bibfield  {title} {\bibinfo {title} {How vortices mix},\ }\href {https://doi.org/10.1017/S0022112002003166} {\bibfield  {journal} {\bibinfo  {journal} {J. Fluid Mech.}\ }\textbf {\bibinfo {volume} {476}},\ \bibinfo {pages} {213} (\bibinfo {year} {2003})}\BibitemShut {NoStop}%
\bibitem [{\citenamefont {Meunier}\ and\ \citenamefont {Villermaux}(2010)}]{Meunier2010}%
  \BibitemOpen
  \bibfield  {author} {\bibinfo {author} {\bibfnamefont {P.}~\bibnamefont {Meunier}}\ and\ \bibinfo {author} {\bibfnamefont {E.}~\bibnamefont {Villermaux}},\ }\bibfield  {title} {\bibinfo {title} {The diffusive strip method for scalar mixing in two dimensions},\ }\href {https://doi.org/10.1017/S0022112010003162} {\bibfield  {journal} {\bibinfo  {journal} {J. Fluid Mech.}\ }\textbf {\bibinfo {volume} {662}},\ \bibinfo {pages} {134} (\bibinfo {year} {2010})}\BibitemShut {NoStop}%
\bibitem [{\citenamefont {Meunier}\ and\ \citenamefont {Villermaux}(2022)}]{Meunier2022}%
  \BibitemOpen
  \bibfield  {author} {\bibinfo {author} {\bibfnamefont {P.}~\bibnamefont {Meunier}}\ and\ \bibinfo {author} {\bibfnamefont {E.}~\bibnamefont {Villermaux}},\ }\bibfield  {title} {\bibinfo {title} {The diffuselet concept for scalar mixing},\ }\href {https://doi.org/10.1017/jfm.2022.771} {\bibfield  {journal} {\bibinfo  {journal} {J. Fluid Mech.}\ }\textbf {\bibinfo {volume} {951}},\ \bibinfo {pages} {A33} (\bibinfo {year} {2022})}\BibitemShut {NoStop}%
\bibitem [{\citenamefont {Kumar}\ \emph {et~al.}(2012)\citenamefont {Kumar}, \citenamefont {Janetzko}, \citenamefont {Schumacher},\ and\ \citenamefont {Shaw}}]{Kumar2012}%
  \BibitemOpen
  \bibfield  {author} {\bibinfo {author} {\bibfnamefont {B.}~\bibnamefont {Kumar}}, \bibinfo {author} {\bibfnamefont {F.}~\bibnamefont {Janetzko}}, \bibinfo {author} {\bibfnamefont {J.}~\bibnamefont {Schumacher}},\ and\ \bibinfo {author} {\bibfnamefont {R.~A.}\ \bibnamefont {Shaw}},\ }\bibfield  {title} {\bibinfo {title} {Extreme responses of a coupled scalar--particle system during turbulent mixing},\ }\href {https://doi.org/10.1088/1367-2630/14/11/115020} {\bibfield  {journal} {\bibinfo  {journal} {New J. Phys.}\ }\textbf {\bibinfo {volume} {14}},\ \bibinfo {pages} {115020} (\bibinfo {year} {2012})}\BibitemShut {NoStop}%
\bibitem [{\citenamefont {Kumar}\ \emph {et~al.}(2013)\citenamefont {Kumar}, \citenamefont {Schumacher},\ and\ \citenamefont {Shaw}}]{Kumar2013}%
  \BibitemOpen
  \bibfield  {author} {\bibinfo {author} {\bibfnamefont {B.}~\bibnamefont {Kumar}}, \bibinfo {author} {\bibfnamefont {J.}~\bibnamefont {Schumacher}},\ and\ \bibinfo {author} {\bibfnamefont {R.~A.}\ \bibnamefont {Shaw}},\ }\bibfield  {title} {\bibinfo {title} {Cloud microphysical effects of turbulent mixing and entrainment},\ }\href {https://doi.org/10.1007/s00162-012-0272-z} {\bibfield  {journal} {\bibinfo  {journal} {Theor. Comput. Fluid Dyn.}\ }\textbf {\bibinfo {volume} {27}},\ \bibinfo {pages} {361} (\bibinfo {year} {2013})}\BibitemShut {NoStop}%
\bibitem [{\citenamefont {Kumar}\ \emph {et~al.}(2014)\citenamefont {Kumar}, \citenamefont {Schumacher},\ and\ \citenamefont {Shaw}}]{Kumar2014}%
  \BibitemOpen
  \bibfield  {author} {\bibinfo {author} {\bibfnamefont {B.}~\bibnamefont {Kumar}}, \bibinfo {author} {\bibfnamefont {J.}~\bibnamefont {Schumacher}},\ and\ \bibinfo {author} {\bibfnamefont {R.~A.}\ \bibnamefont {Shaw}},\ }\bibfield  {title} {\bibinfo {title} {Lagrangian mixing dynamics at the cloudy–clear air interface},\ }\href {https://doi.org/10.1175/JAS-D-13-0294.1} {\bibfield  {journal} {\bibinfo  {journal} {J. Atmos. Sci.}\ }\textbf {\bibinfo {volume} {71}},\ \bibinfo {pages} {2564} (\bibinfo {year} {2014})}\BibitemShut {NoStop}%
\bibitem [{\citenamefont {G{\"o}tzfried}\ \emph {et~al.}(2017)\citenamefont {G{\"o}tzfried}, \citenamefont {Kumar}, \citenamefont {Shaw},\ and\ \citenamefont {Schumacher}}]{Goetzfried2017}%
  \BibitemOpen
  \bibfield  {author} {\bibinfo {author} {\bibfnamefont {P.}~\bibnamefont {G{\"o}tzfried}}, \bibinfo {author} {\bibfnamefont {B.}~\bibnamefont {Kumar}}, \bibinfo {author} {\bibfnamefont {R.~A.}\ \bibnamefont {Shaw}},\ and\ \bibinfo {author} {\bibfnamefont {J.}~\bibnamefont {Schumacher}},\ }\bibfield  {title} {\bibinfo {title} {{Droplet dynamics and fine-scale structure in a shearless turbulent mixing layer with phase changes}},\ }\href {https://doi.org/10.1017/jfm.2017.23} {\bibfield  {journal} {\bibinfo  {journal} {J. Fluid Mech.}\ }\textbf {\bibinfo {volume} {814}},\ \bibinfo {pages} {452} (\bibinfo {year} {2017})}\BibitemShut {NoStop}%
\bibitem [{\citenamefont {Celani}\ \emph {et~al.}(2005)\citenamefont {Celani}, \citenamefont {Falkovich}, \citenamefont {Mazzino},\ and\ \citenamefont {Seminara}}]{Celani2005}%
  \BibitemOpen
  \bibfield  {author} {\bibinfo {author} {\bibfnamefont {A.}~\bibnamefont {Celani}}, \bibinfo {author} {\bibfnamefont {G.}~\bibnamefont {Falkovich}}, \bibinfo {author} {\bibfnamefont {A.}~\bibnamefont {Mazzino}},\ and\ \bibinfo {author} {\bibfnamefont {A.}~\bibnamefont {Seminara}},\ }\bibfield  {title} {\bibinfo {title} {Droplet condensation in turbulent flows},\ }\href {https://doi.org/10.1209/epl/i2005-10040-4} {\bibfield  {journal} {\bibinfo  {journal} {Europhys. Lett.}\ }\textbf {\bibinfo {volume} {70}},\ \bibinfo {pages} {775} (\bibinfo {year} {2005})}\BibitemShut {NoStop}%
\bibitem [{\citenamefont {Pushenko}\ and\ \citenamefont {Schumacher}(2024)}]{Pushenko2024}%
  \BibitemOpen
  \bibfield  {author} {\bibinfo {author} {\bibfnamefont {V.}~\bibnamefont {Pushenko}}\ and\ \bibinfo {author} {\bibfnamefont {J.}~\bibnamefont {Schumacher}},\ }\bibfield  {title} {\bibinfo {title} {{Connecting finite-time Lyapunov exponents with supersaturation and droplet dynamics in a turbulent bulk flow}},\ }\href {https://doi.org/10.1103/PhysRevE.109.045101} {\bibfield  {journal} {\bibinfo  {journal} {Phys. Rev. E}\ }\textbf {\bibinfo {volume} {109}},\ \bibinfo {pages} {045101} (\bibinfo {year} {2024})}\BibitemShut {NoStop}%
\bibitem [{\citenamefont {Schumacher}(2007)}]{Schumacher2007}%
  \BibitemOpen
  \bibfield  {author} {\bibinfo {author} {\bibfnamefont {J.}~\bibnamefont {Schumacher}},\ }\bibfield  {title} {\bibinfo {title} {{Sub-Kolmogorov-scale fluctuations in fluid turbulence}},\ }\href {https://doi.org/10.1209/0295-5075/80/54001} {\bibfield  {journal} {\bibinfo  {journal} {Europhys. Lett.}\ }\textbf {\bibinfo {volume} {80}},\ \bibinfo {pages} {54001} (\bibinfo {year} {2007})}\BibitemShut {NoStop}%
\bibitem [{\citenamefont {{de Rivas, A. and Villermaux, E.}}(2016)}]{Rivas2016}%
  \BibitemOpen
  \bibfield  {author} {\bibinfo {author} {\bibnamefont {{de Rivas, A. and Villermaux, E.}}},\ }\bibfield  {title} {\bibinfo {title} {Dense spray evaporation as a mixing process},\ }\href {https://doi.org/10.1103/PhysRevFluids.1.014201} {\bibfield  {journal} {\bibinfo  {journal} {Phys. Rev. Fluids}\ }\textbf {\bibinfo {volume} {1}},\ \bibinfo {pages} {014201} (\bibinfo {year} {2016})}\BibitemShut {NoStop}%
\bibitem [{\citenamefont {Kolmogorov}(1941)}]{Kolmogorov1941}%
  \BibitemOpen
  \bibfield  {author} {\bibinfo {author} {\bibfnamefont {A.~N.}\ \bibnamefont {Kolmogorov}},\ }\bibfield  {title} {\bibinfo {title} {{The local structure of turbulence in incompressible viscous fluid for very large Reynolds numbers}},\ }\href@noop {} {\bibfield  {journal} {\bibinfo  {journal} {Dokl. Akad. Nauk SSSR}\ }\textbf {\bibinfo {volume} {30}},\ \bibinfo {pages} {9} (\bibinfo {year} {1941})}\BibitemShut {NoStop}%
\bibitem [{\citenamefont {Obukhov}(1949)}]{Obukhov1949}%
  \BibitemOpen
  \bibfield  {author} {\bibinfo {author} {\bibfnamefont {A.~M.}\ \bibnamefont {Obukhov}},\ }\bibfield  {title} {\bibinfo {title} {Structure of the temperature field in a turbulent flow},\ }\href@noop {} {\bibfield  {journal} {\bibinfo  {journal} {Izv. Akad. Nauk SSSR, Ser. Geogr. Geofiz.}\ }\textbf {\bibinfo {volume} {13}},\ \bibinfo {pages} {58} (\bibinfo {year} {1949})}\BibitemShut {NoStop}%
\bibitem [{\citenamefont {Corrsin}(1951)}]{Corrsin1951}%
  \BibitemOpen
  \bibfield  {author} {\bibinfo {author} {\bibfnamefont {S.}~\bibnamefont {Corrsin}},\ }\bibfield  {title} {\bibinfo {title} {On the spectrum of isotropic temperature fluctuations in an isotropic turbulence},\ }\href {https://doi.org/10.1063/1.1699986} {\bibfield  {journal} {\bibinfo  {journal} {J. Appl. Phys.}\ }\textbf {\bibinfo {volume} {22}},\ \bibinfo {pages} {469} (\bibinfo {year} {1951})}\BibitemShut {NoStop}%
\bibitem [{\citenamefont {Ottino}(1990)}]{Ottino1990}%
  \BibitemOpen
  \bibfield  {author} {\bibinfo {author} {\bibfnamefont {J.~M.}\ \bibnamefont {Ottino}},\ }\bibfield  {title} {\bibinfo {title} {Mixing, chaotic advection, and turbulence},\ }\href {https://doi.org/10.1146/annurev.fl.22.010190.001231} {\bibfield  {journal} {\bibinfo  {journal} {Annu. Rev. Fluid Mech.}\ }\textbf {\bibinfo {volume} {22}},\ \bibinfo {pages} {207} (\bibinfo {year} {1990})}\BibitemShut {NoStop}%
\bibitem [{\citenamefont {Pekurovsky}(2012)}]{Pekurovsky2012}%
  \BibitemOpen
  \bibfield  {author} {\bibinfo {author} {\bibfnamefont {D.}~\bibnamefont {Pekurovsky}},\ }\bibfield  {title} {\bibinfo {title} {{P3DFFT: A framework for parallel computations of Fourier transforms in three dimensions}},\ }\href {https://doi.org/10.1137/11082748X} {\bibfield  {journal} {\bibinfo  {journal} {SIAM J. Sci. Comput.}\ }\textbf {\bibinfo {volume} {34}},\ \bibinfo {pages} {C192} (\bibinfo {year} {2012})}\BibitemShut {NoStop}%
\bibitem [{\citenamefont {Yeung}\ and\ \citenamefont {Pope}(1988)}]{Yeung1988}%
  \BibitemOpen
  \bibfield  {author} {\bibinfo {author} {\bibfnamefont {P.~K.}\ \bibnamefont {Yeung}}\ and\ \bibinfo {author} {\bibfnamefont {S.~B.}\ \bibnamefont {Pope}},\ }\bibfield  {title} {\bibinfo {title} {{An algorithm for tracking fluid particles in numerical simulations of homogeneous turbulence}},\ }\href {https://doi.org/10.1016/0021-9991(88)90022-8} {\bibfield  {journal} {\bibinfo  {journal} {J. Comp. Phys.}\ }\textbf {\bibinfo {volume} {79}},\ \bibinfo {pages} {373} (\bibinfo {year} {1988})}\BibitemShut {NoStop}%
\bibitem [{\citenamefont {Ranz}(1979)}]{Ranz1979}%
  \BibitemOpen
  \bibfield  {author} {\bibinfo {author} {\bibfnamefont {W.~E.}\ \bibnamefont {Ranz}},\ }\bibfield  {title} {\bibinfo {title} {Applications of a stretch model to mixing, diffusion, and reaction in laminar and turbulent flows},\ }\href {https://doi.org/10.1002/aic.690250105} {\bibfield  {journal} {\bibinfo  {journal} {AIChE J.}\ }\textbf {\bibinfo {volume} {25}},\ \bibinfo {pages} {41} (\bibinfo {year} {1979})}\BibitemShut {NoStop}%
\bibitem [{\citenamefont {Ashurst}\ \emph {et~al.}(1987)\citenamefont {Ashurst}, \citenamefont {Kerstein}, \citenamefont {Kerr},\ and\ \citenamefont {Gibson}}]{Ashurst1987}%
  \BibitemOpen
  \bibfield  {author} {\bibinfo {author} {\bibfnamefont {W.~T.}\ \bibnamefont {Ashurst}}, \bibinfo {author} {\bibfnamefont {A.~R.}\ \bibnamefont {Kerstein}}, \bibinfo {author} {\bibfnamefont {R.~M.}\ \bibnamefont {Kerr}},\ and\ \bibinfo {author} {\bibfnamefont {C.~H.}\ \bibnamefont {Gibson}},\ }\bibfield  {title} {\bibinfo {title} {{Alignment of vorticity and scalar gradient with strain rate in simulated Navier–Stokes turbulence}},\ }\href {https://doi.org/10.1063/1.866513} {\bibfield  {journal} {\bibinfo  {journal} {Phys. Fluids}\ }\textbf {\bibinfo {volume} {30}},\ \bibinfo {pages} {2343} (\bibinfo {year} {1987})}\BibitemShut {NoStop}%
\bibitem [{\citenamefont {Girimaji}\ and\ \citenamefont {Pope}(1990)}]{Girimaji1990}%
  \BibitemOpen
  \bibfield  {author} {\bibinfo {author} {\bibfnamefont {S.~S.}\ \bibnamefont {Girimaji}}\ and\ \bibinfo {author} {\bibfnamefont {S.~B.}\ \bibnamefont {Pope}},\ }\bibfield  {title} {\bibinfo {title} {Material-element deformation in isotropic turbulence},\ }\href {https://doi.org/10.1017/S0022112090003330} {\bibfield  {journal} {\bibinfo  {journal} {J. Fluid Mech.}\ }\textbf {\bibinfo {volume} {220}},\ \bibinfo {pages} {427} (\bibinfo {year} {1990})}\BibitemShut {NoStop}%
\bibitem [{\citenamefont {Zinchenko}\ \emph {et~al.}(2024)\citenamefont {Zinchenko}, \citenamefont {Pushenko},\ and\ \citenamefont {Schumacher}}]{Zinchenko2024}%
  \BibitemOpen
  \bibfield  {author} {\bibinfo {author} {\bibfnamefont {G.}~\bibnamefont {Zinchenko}}, \bibinfo {author} {\bibfnamefont {V.}~\bibnamefont {Pushenko}},\ and\ \bibinfo {author} {\bibfnamefont {J.}~\bibnamefont {Schumacher}},\ }\bibfield  {title} {\bibinfo {title} {{Local precursors to anomalous dissipation in Navier-Stokes turbulence: Burgers vortex-type models and simulation analysis}},\ }\href {https://doi.org/10.1103/PhysRevFluids.9.114608} {\bibfield  {journal} {\bibinfo  {journal} {Phys. Rev. Fluids}\ }\textbf {\bibinfo {volume} {9}},\ \bibinfo {pages} {114608} (\bibinfo {year} {2024})}\BibitemShut {NoStop}%
\bibitem [{\citenamefont {Burton}\ and\ \citenamefont {Dahm}(2005)}]{Burton2005}%
  \BibitemOpen
  \bibfield  {author} {\bibinfo {author} {\bibfnamefont {G.~C.}\ \bibnamefont {Burton}}\ and\ \bibinfo {author} {\bibfnamefont {W.~J.~A.}\ \bibnamefont {Dahm}},\ }\bibfield  {title} {\bibinfo {title} {{Multifractal subgrid-scale modeling for large-eddy simulation. II. Backscatter limiting and \textit{a posteriori} evaluation}},\ }\href {https://doi.org/10.1063/1.1965094} {\bibfield  {journal} {\bibinfo  {journal} {Phys. Fluids}\ }\textbf {\bibinfo {volume} {17}},\ \bibinfo {pages} {075112} (\bibinfo {year} {2005})}\BibitemShut {NoStop}%
\bibitem [{\citenamefont {Burton}(2008)}]{Burton2008}%
  \BibitemOpen
  \bibfield  {author} {\bibinfo {author} {\bibfnamefont {G.~C.}\ \bibnamefont {Burton}},\ }\bibfield  {title} {\bibinfo {title} {{The nonlinear large-eddy simulation method applied to $Sc\approx 1$ and $Sc\gg 1$ passive-scalar mixing}},\ }\href {https://doi.org/10.1063/1.2840199} {\bibfield  {journal} {\bibinfo  {journal} {Phys. Fluids}\ }\textbf {\bibinfo {volume} {20}},\ \bibinfo {pages} {035103} (\bibinfo {year} {2008})}\BibitemShut {NoStop}%
\bibitem [{\citenamefont {G{\"o}tzfried}\ \emph {et~al.}(2019)\citenamefont {G{\"o}tzfried}, \citenamefont {Emran}, \citenamefont {Villermaux},\ and\ \citenamefont {Schumacher}}]{Goetzfried2019}%
  \BibitemOpen
  \bibfield  {author} {\bibinfo {author} {\bibfnamefont {P.}~\bibnamefont {G{\"o}tzfried}}, \bibinfo {author} {\bibfnamefont {M.~S.}\ \bibnamefont {Emran}}, \bibinfo {author} {\bibfnamefont {E.}~\bibnamefont {Villermaux}},\ and\ \bibinfo {author} {\bibfnamefont {J.}~\bibnamefont {Schumacher}},\ }\bibfield  {title} {\bibinfo {title} {{Comparison of Lagrangian and Eulerian frames of passive scalar turbulent mixing}},\ }\href {https://doi.org/10.1103/PhysRevFluids.4.044607} {\bibfield  {journal} {\bibinfo  {journal} {Phys. Rev, Fluids}\ }\textbf {\bibinfo {volume} {4}},\ \bibinfo {pages} {044607} (\bibinfo {year} {2019})}\BibitemShut {NoStop}%
\bibitem [{\citenamefont {Srivastava}(1989)}]{Srivastava1989}%
  \BibitemOpen
  \bibfield  {author} {\bibinfo {author} {\bibfnamefont {R.~C.}\ \bibnamefont {Srivastava}},\ }\bibfield  {title} {\bibinfo {title} {{Growth of cloud drops by condensation: A criticism of currently accepted theory and a new approach}},\ }\href {https://doi.org/10.1175/1520-0469(1989)046<0869:GOCDBC>2.0.CO;2} {\bibfield  {journal} {\bibinfo  {journal} {J. Atmos. Sci.}\ }\textbf {\bibinfo {volume} {46}},\ \bibinfo {pages} {869} (\bibinfo {year} {1989})}\BibitemShut {NoStop}%
\bibitem [{\citenamefont {Villermaux}\ \emph {et~al.}(2017)\citenamefont {Villermaux}, \citenamefont {Moutte}, \citenamefont {Amielh},\ and\ \citenamefont {Meunier}}]{Villermaux2017}%
  \BibitemOpen
  \bibfield  {author} {\bibinfo {author} {\bibfnamefont {E.}~\bibnamefont {Villermaux}}, \bibinfo {author} {\bibfnamefont {A.}~\bibnamefont {Moutte}}, \bibinfo {author} {\bibfnamefont {M.}~\bibnamefont {Amielh}},\ and\ \bibinfo {author} {\bibfnamefont {P.}~\bibnamefont {Meunier}},\ }\bibfield  {title} {\bibinfo {title} {Fine structure of the vapor field in evaporating dense sprays},\ }\href {https://doi.org/10.1103/PhysRevFluids.2.074501} {\bibfield  {journal} {\bibinfo  {journal} {Phys. Rev. Fluids}\ }\textbf {\bibinfo {volume} {2}},\ \bibinfo {pages} {074501} (\bibinfo {year} {2017})}\BibitemShut {NoStop}%
\end{thebibliography}%

\end{document}